\documentclass{elsarticle}
\usepackage{xcolor}
\usepackage{graphicx}
\usepackage[utf8]{inputenc}
\usepackage{amsmath}
\usepackage{xcolor}
\usepackage[margin=3cm]{geometry}
\usepackage{setspace}

\usepackage{hyperref}

\setcitestyle{authoryear,round,semicolon}
\makeatletter
\providecommand{\doi}[1]{%
	\begingroup
	\let\bibinfo\@secondoftwo
	\urlstyle{rm}%
	\href{http://dx.doi.org/#1}{%
		doi:\discretionary{}{}{}%
		\nolinkurl{#1}%
	}%
	\endgroup
}
\makeatother

\usepackage{cleveref}

\newcommand{\pdts}[1]{\partial_t #1}
\newcommand{\pdt}[1]{\partial_t\left(#1\right)}
\newcommand{\pdxs}[1]{\partial_x #1}
\newcommand{\pdx}[1]{\partial_x\left(#1\right)}
\newcommand{\pdis}[1]{\partial_i #1}

\newcommand{\pdjs}[1]{\partial_j #1}
\newcommand{\pdj}[1]{\partial_j\left(#1\right)}
\newcommand{\pdks}[1]{\partial_k #1}
\newcommand{\pdk}[1]{\partial_k\left(#1\right)}

\newcommand{\vavg}[1]{\overline{#1}}
\newcommand{\phavg}[1]{\langle #1 \rangle}
\newcommand{\favg}[1]{\tilde{#1}}
\newcommand{\wfavg}[1]{\widetilde{#1}}

\newcommand{\fu}{\favg{u}}
\newcommand{\phrho}{\phavg{\rho}}

\newcommand{\php}{\phavg{p}}

\makeatletter
\def\ps@pprintTitle{%
	\let\@oddhead\@empty
	\let\@evenhead\@empty
	\def\@oddfoot{}%
	\let\@evenfoot\@oddfoot}
\makeatother

\usepackage{etoolbox}
\patchcmd{\pprintMaketitle}
{\hrule\vskip12pt}
{\hrule\vskip12pt\ifvoid\extrainfobox\else\unvbox\extrainfobox\par\vskip12pt\fi}
{}{}

\newenvironment{extrainfo}
{\global\setbox\extrainfobox=\vbox\bgroup\parindent=0pt }
{\egroup}
\newsavebox\extrainfobox

\begin{document}
\begin{frontmatter}
	
\title{Computational analysis of shock-induced flow through stationary particle clouds}




\author[a]{Andreas Nygård Osnes}
\ead{a.n.osnes@its.uio.no}
\address[a]{Department of Technology Systems, University of Oslo, PO Box 70 Kjeller, 2027 Kjeller, Norway.}
\author[b]{Magnus Vartdal}
\ead{Magnus.Vartdal@ffi.no}
\address[b]{Norwegian Defence Research Establishment, PO Box 25 Kjeller, 2027 Kjeller, Norway}
\author[c,d]{Marianne Gjestvold Omang}
\ead{m.g.omang@astro.uio.no}
\address[c]{Norwegian Defence Estates Agency, PO Box 405 Sentrum, 0103 Oslo, Norway.}
\address[d]{Institude of Theoretical Astrophysics, University of Oslo, PO Box 1029 Blindern, 0315 Oslo, Norway.}
\author[a]{Bjørn Anders Pettersson Reif}
\ead{b.a.p.reif@its.uio.no}

\begin{extrainfo}
This is the author's accepted manuscript of the original manuscript published at 
	
\noindent\url{https://doi.org/10.1016/j.ijmultiphaseflow.2019.03.010}

\textcopyright 2019. This manuscript version is made available under the CC-BY-NC-ND 4.0 license \url{https://creativecommons.org/licenses/by-nc-nd/4.0/}
\end{extrainfo}

\begin{abstract}

\noindent We investigate the shock-induced flow through random particle arrays using particle-resolved Large Eddy Simulations for different incident shock wave Mach numbers, particle volume fractions and particle sizes. We analyze trends in mean flow quantities and the unresolved terms in the volume averaged momentum equation, as we vary the three parameters. We find that the shock wave attenuation and certain mean flow trends can be predicted by the opacity of the particle cloud, which is a function of particle size and particle volume fraction. We show that the Reynolds stress field plays an important role in the momentum balance at the particle cloud edges, and therefore strongly affects the reflected shock wave strength. The Reynolds stress was found to be insensitive to particle size, but strongly dependent on particle volume fraction. It is in better agreement with results from simulations of flow through particle clouds at fixed mean slip Reynolds numbers in the incompressible regime, than with results from other shock wave particle cloud studies, which have utilized either inviscid or two-dimensional approaches. We propose an algebraic model for the streamwise Reynolds stress based on the observation that the separated flow regions are the primary contributions to the Reynolds stress.
\end{abstract}

\end{frontmatter}

\section{Introduction}

Interaction between shock waves and particle clouds are of interest in a number of different natural phenomena, as well as industrial applications and safety measures such as shock wave mitigation using porous barriers \citep{suzuki2000,chaudhuri2013}. It also finds applications in heterogeneous explosives \citep{zhang2006} and explosive dissemination of powders and liquids \citep{zhang2001,milne2010,rodriguez2017}. In coal mines, enhanced or secondary explosions due to coal dust is a major safety concern \citep{ugarte2017,shimura2018}. Shock wave particle interaction also occurs in a number of natural phenomena, with volcanic eruptions \citep{bower1996} as the prime example. There are also astrophysical examples such as ejection of stellar dust from supernovae \citep{silvia2012}. More generally, high-speed multiphase flow has important industrial applications, such as liquid and solid fuel engines and fluidized beds. Gas-turbines operating in regions with suspensions of sand particles in the air are subject to substantial degradation due to particle deposition on turbine blades \citep{hamed2006}. Water injection systems have been used to reduce sound intensity at rocket launch pads \citep{ignatius2008}, and it might be possible to utilize similar systems to reduce jet noise \citep{krothapalli2003}, which is especially important around air-crafts during take-off. 

Shock wave interaction with particle clouds has been extensively studied over the last fifty years. The dilute particle cloud and the granular flow regimes are quite well understood, but the intermediate regime has proven challenging to model \citep{theofanous2017-2}. The intermediate volume fraction regime is where particles neither display the same statistical properties as isolated particles, nor as particles in the granular regime. In \citet{crowe2011}, flows with particle volume fractions above $0.1\%$ were considered to belong to this regime, while \citet{zhang2001} used a lower limit of $1\%$. The difficulty in modeling these flows stems from the complex interaction between the flow field and the particle distribution. The particles occupy a volume that is large enough that their collective nozzle effect is one of the dominanting dynamical effects in the flow \citep{mehta2018}, but they are not close enough that particle collision exclusively determines the movement of the particles. Each particle deflects the flow around it, causing local flow acceleration and deceleration that depends on the local particle configuration. Additionally, boundary layers develop over the particle surface, and the flow separates behind the particle. When the shock wave passes over a particle there is a reflection from the front of the particle, and a focusing of the shock wave behind it \citep{tanno2003}. The reflected shock interacts with the upstream particles and their wakes, and also with reflected shocks from other nearby particles. These complex flow dynamics lead to a large variation in drag forces that depends on the local particle configuration.

The intermediate particle volume fraction regime has recently become feasible to study in much greater detail than was previously possible. In experiments, the short time-scales and the limited possibility of recording data in the regions of interest have presented significant difficulties. Recent improvements to experimental techniques have enabled experimentalists to accurately characterize the wave system and particle distribution when a shock wave passes through a curtain of particles \citep{Wagner2012,ling2012}. Even more recently, \citet{demauro2017} used high speed, time-resolved, particle image velocimetry to measure velocity fields in front of and behind the particle curtain. The new sets of experimental data have resulted in a renewed effort to study these problems using numerical simulations, in particular using the Eulerian-Lagrangian framework \citep{houim2016,shallcross2018,theofanous2017-2}, but also Eulerian-Eulerian models \citep{mcgrath2016,saurel2017,utkin2017}. 

Some quantities are very difficult to measure experimentally, such as flow field distributions inside the particle cloud and fluctuations at the particle scale. Instead, these can be obtained using particle-resolved numerical simulations. Such simulations are computationally expensive, since a large number of particles must be used to obtain meaningful statistics. A limiting factor is the very large scale separation between the dynamically important particle scale physics and the global length scale of the problems. However, a number of studies have recently been able to investigate shock-wave particle cloud interaction using two-dimensional \citep{regele2014,hosseinzadeh2018,sen2018} and three-dimensional \citep{sridharan2015,mehta2016,mehta2018-b,mehta2018,theofanous2018} numerical simulations. In particular, particle resolved simulations can be used to investigate closures for unresolved terms that appear in Eulerian-Lagrangian or Eulerian-Eulerian models due to averaging of products of fluctuations. Volume averaging is one form of filter used in Large Eddy Simulations, and is also used in the formulation of most Eulerian-Lagrangian methods. Volume averaging does not commute with spatial or temporal derivatives. Therefore, the averaging operation introduces new terms in the volume averaged equations, as discussed in e.g. \citet{schwarzkopf2015}. Volume averaging can be applied to data from particle-resolved simulations to investigate the unclosed terms, both the terms that appear in single-phase models and those that are specific to dispersed flow models. 

In the intermediate particle volume fraction regime, the inter-particle distance and the particle size are of the same order. The spatial extent of the flow field fluctuations is comparable to the inter-particle distance, and we will refer to these as particle scale fluctuations. It is common to divide the flow fluctuations into pseudo-turbulent and turbulent structures, and refer to the kinetic energy in these as pseudo-turbulent kinetic energy (PTKE) and turbulent kinetic energy (TKE), respectively. The flow perturbations induced by the particles are considered to be pseudo-turbulent effects. Pseudo-turbulent flow structures might have very different time and spatial scales than turbulent structures. For this reason it is a useful technique for analyses and modeling purposes to distinguish between the two \citep{mehrabadi2015}. In the setting of shock waves passing through particle clouds, PTKE is caused primarily by three effects. Firstly, shock wave reflection from individual particles causes very large differences between the region affected by the reflected shock and the surrounding regions, with correspondingly high PTKE values. Secondly, flow deflection around particles causes local flow accelerations and decelerations, resulting in both streamwise and spanwise fluctuations. In addition, flow separation behind particles also causes a significant deviation from the mean flow speed. This last effect in particular will be discussed in this paper. As evident from these examples, pseudo-turbulent flow fluctuations are quite different from classical turbulence. However, pseudo-turbulent fluctuations might themselves generate classical turbulent fluctuations. This is expected to occur as a result of the strong velocity shear in the particle wakes. The processes generating TKE and PTKE are very different phenomena, and should therefore be modeled differently. Both TKE and PTKE enter the volume averaged momentum equations through a term that is analogous to the classical Reynolds stress, and for convenience we will use the term Reynolds stress for this term throughout this paper. 

The velocity fluctuations in shock-particle interaction have previously been examined in two-dimensional flows using both inviscid \citep{regele2014} and viscous simulations \citep{hosseinzadeh2018}. In those flows, the PTKE was found to be slightly higher in the inviscid simulations, but of the same order as the mean flow kinetic energy in both cases. \citet{regele2014} additionally demonstrated the importance of capturing the PTKE in volume-averaged simulations in order to obtain correct pressure fields. In contrast, \citet{mehta2018-b} found very low values of PTKE, demonstrating a significant difference between the two-dimensional and inviscid three-dimensional simulations. The Reynolds stress plays an important role in the dynamics around the particle cloud edges, and in particular it influences the (time-dependent) strength of the reflected shock wave. Since the incoming flow field is altered by the reflected shock wave, phenomena such as particle drag, pressure drop through the particle cloud and even the transmitted shock strength depend directly on the strength of the reflected shock, and therefore also on the Reynolds stress. 	

Recent studies have recognized that an issue for Eulerian-Lagrangian methods is that the forces imposed on the continuous phase by a particle disturbs the flow around the particle. Calculation of drag by standard drag laws is incorrect if continuous phase variables within the disturbed flow region is used, because most drag-laws are calibrated against undisturbed flow quantities. Methods for handling this problem exactly in the zero Reynolds number limit have been proposed and even shown to yield good results in finite Reynolds number flows \citep{horwitz2016,horwitz2018,balachandar2019}. Accounting for how the particle influences the local flow field was also done in \cite{moore2018}, who used a linear superposable wake to approximate continuous phase fluctuations within a particle cloud for incompressible flow. So far, no such model has been proposed for the compressible flow inside a particle cloud. However, using knowledge about how the particles disturb the flow in their vicinity to improve drag computation can be done even in this complex setting, as will be shown. 

In this work we perform three-dimensional, time dependent, viscous simulations of a shock wave passing through a random particle array. The particles are assumed to be inert and stationary. We vary the incident shock wave Mach number between $2.2$ and $3$, the particle volume fraction between 0.05 and 0.1, and the particle diameter between $50$ $\mu$m and $100$ $\mu$m. Preliminary results from these simulations were reported in \cite{vartdal2018}. We utilize volume averaging to define the mean flow and the fluctuations from that mean. Key flow properties such as mean velocity, pressure and density, as well as the Reynolds stress and its anisotropy, are examined.

The purpose of this work is twofold. Firstly, we analyze trends in mean flow properties over different combinations of volume fractions and Mach numbers than those that have been reported previously. In addition, we vary the particle diameter, which has not been done before. This analysis improves the understanding of the bulk effect of particle cloud properties on shock wave particle cloud interaction. The second purpose is to analyze statistics of the Reynolds stresses and their anisotropy, as well as their importance in the flow dynamics. To the authors' knowledge, this has not yet been investigated for viscous simulations of three-dimensional random particle arrays. Such data are crucial to the development of Reynolds stress closure models for shock wave particle cloud interaction. 

This paper is organized as follows. In \cref{sec:shockinducedflow} we briefly introduce the basic flow patterns occurring when the shock wave passes over a group of particles. The governing equations and the volume averaged equations used for analysis are described in \cref{sec:goveq}. \Cref{sec:comp-set-up} describes the computational method and the set-up of the problems under consideration. \Cref{sec:results} presents the simulation results. First the grid-quality is checked by examination of particle drag and resolution of viscous shear layers. Next, we discuss the shock wave attenuation as it passes through the particle layer. We then examine the mean fields, i.e. density, velocity, pressure and Mach number distributions throughout the particle layer, and discuss trends as we change particle volume fractions, particle diameters and incident shock wave Mach numbers. Next we discuss the velocity fluctuations and their anisotropy. We discuss the momentum balance around the upstream edge of the particle cloud, to highlight the dynamic importance of the velocity fluctuations. The discussion of the simulation results is finalized by an examination of the particle drag coefficients and the average particle forces obtained in the different simulations. In \cref{sec:modeling} we utilize the data from the resolved simulations to propose models that capture some of the observed properties of the flow. We provide an algebraic expression for combinations of particle diameter and particle volume fractions that result in the same shock wave attenuation. We also propose an algebraic Reynolds stress model based on the effect of separated flow behind particles on volume averaged equations, and compare this model to the streamwise velocity fluctuation intensity obtained in the simulations. Finally, concluding remarks are given in \cref{sec:conclusions}.

\section{Shock-induced flow around particles}
\label{sec:shockinducedflow}

\begin{figure*}
	\centerline{
	\includegraphics[scale=1.0]{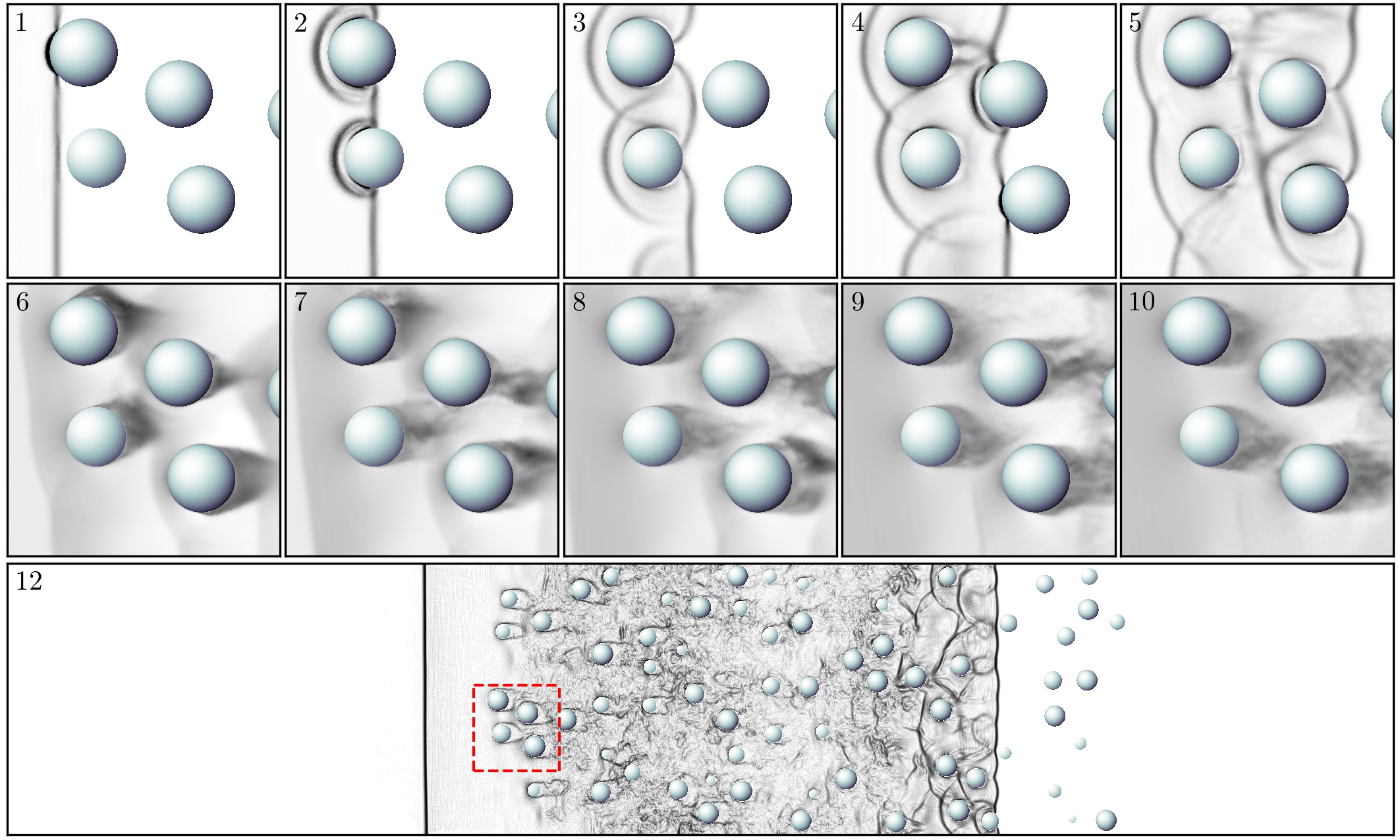}}
	\caption{Numerical schlieren images (top and bottom row) and streamwise velocities (middle row) in a cut plane, when a shock wave impacts on and passes through a cloud of particles. Flow direction is from left to right. Frames are taken at times $(t-t_0)/\tau_\mathrm{p}= 0.13,\ 0.69,\ 1.25,\ 1.79,\ 2.86,\ 4.81,\ 6.77,\ 8.73,\ 10.63,\ 12.47,\ 26.82$. $t_0$ denotes the time when the shock wave is at $x=0$ (upstream particle cloud edge) and $\tau_\mathrm{p}$ is defined in \cref{eq:timescales}. In the middle frames, the colormap is linear between $-900$ m/s (black) and $900$ m/s (white). The red dashed square shows the location of the zoomed view in the upper rows.}
	\label{fig:flowsnapshots}
\end{figure*}

In this section, we present a brief overview of the flow during and after the shock wave passes over a group of particles. In \cref{fig:flowsnapshots}, numerical schlieren images \citep{quirk1997} and instantaneous streamwise velocities are shown for a time-series in one of the simulations (case VII) that will be described in \cref{sec:comp-set-up}. The time sequence covers the shock wave pattern and the subsequent development of particle wakes.

Initially, a planar shock impacts on the first particle, and a regular shock reflection is formed at the front of the particle (first frame). As the shock propagates, a Mach reflection is obtained, which can be discerned in the second frame. During this time, the pressure difference between the front and the back of the particle is very large. Due to the presence of multiple particles, the individual reflected shocks coalesce and form a reflected shock \citep{boiko1997,Wagner2012}, which over time becomes nearly planar and propagates upstream. The particles also cause shock wave diffraction, as clearly seen by the curved front in the third frame. Behind each particle, the shock is focused and a high-pressure region is created (fourth frame).
	
Viscous forces become more important for the flow around particles when the particle Reynolds number is reduced. This is relevant also for shock particle interaction, since smaller particles correspond to lower particle Reynolds numbers. Henceforth, "particle Reynolds number" and "Reynolds number" will be used interchangeably. \citet{sun2005} showed that depending on the Reynolds number, the high-pressure region created by shock focusing can create temporary negative drag-coefficients. This phenomena was only observed for particle Reynolds numbers of the order $10^3$. For lower Reynolds numbers, viscous forces counteracted this effect, and the total drag-coefficient remained positive. As they varied the particle Reynolds number from 4900 to 49, the importance of viscous forces increased drastically, and for the lower Reynolds number, the late-time viscous forces was almost twice the magnitude of the pressure forces.

The particle Reynolds number is often used to characterize the flow, and it is typically based on undisturbed flow, or incident flow, quantities. It is likely that characterization based on incident flow properties are less reliable in the shock wave particle cloud setting, due to the generation of a collective reflected shock wave. The strength of this shock wave determines the properties of the gas that enters the particle cloud, and is highly dependent on properties of the cloud. In addition, particles within the cloud are exposed to the pseudo-turbulent flow induced by upstream particles, which is very likely to affect statistical properties of the wake. 

The development of flow separation and particle wakes can be seen in the middle frames in \cref{fig:flowsnapshots}. Boundary layers develop over the particle surfaces, and the flow separates. The particle wakes are highly distorted due to the presence of other particles, even for the first particles in the cloud. For isolated particles with Reynolds numbers in the range $50-300$, \citet{nagata2016} found that for increasing Reynolds numbers, the separation line moves forward along the particle surface. They also found that higher Reynolds numbers resulted in longer separated flow regions. Within a particle cloud, the length of the separated flow region is significantly affected by the presence of other particles, as is the case for the wake behind the leftmost particle in \cref{fig:flowsnapshots}. This phenomena cannot be described solely based on Reynolds number and Mach numbers.

The bottom frame of \cref{fig:flowsnapshots} shows a snapshot of the flow over the whole particle cloud. The different effects discussed above are visible in this frame, occurring at different spatial locations throughout the particle cloud.

\section{Governing equations}	
\label{sec:goveq}
The gas-dynamic processes considered in this work are governed by the conservation equations of mass, momentum and energy. In differential form these are
\begin{equation}
\pdts{\rho}+\pdk{\rho u_k}=0,
\label{eq:mass}
\end{equation}
\begin{equation}								
\pdt{\rho u_i}+\pdk{\rho u_i u_k}=-\pdis{p} + \pdjs{\sigma_{ij}},
\label{eq:momentum}
\end{equation}
\begin{equation}
\pdt{\rho E}+\pdk{\rho E u_k + p u_k}=\pdj{\sigma_{ij}u_i} - \pdk{\lambda\pdks{T}},
\label{eq:energy}
\end{equation}
where $\rho$ is the mass density, $u$ is the velocity, $p$ is the pressure, $\sigma_{ij}=\mu (\pdjs{u_i}+\pdis{u_j}-2\pdks{u_k}\delta_{ij}/3)$ is the viscous stress tensor,  $\mu$ is the dynamic viscosity, $E=\rho e + 0.5\rho u_ku_k$ is the total energy per unit volume, $e$ is the internal energy per unit mass, $\lambda$ is the thermal conductivity, and $T$ is the temperature. We utilize the ideal gas equation of state, with $\gamma=1.4$, and relate internal energy and temperature by a constant specific heat capacity. We assume a power law dependence of viscosity on temperature, with an exponent of $0.76$, and we relate the thermal diffusivity to the viscosity by assuming a constant Prandtl number of $0.7$. 

In the analysis of the results, we consider the volume averaged equations of motion. These are obtained by applying the volume averaging operator to \cref{eq:mass,eq:momentum,eq:energy}. We use the notation $\vavg{\cdot}$ for volume averaging, $\langle\cdot\rangle$ for phase-averaging, and $\favg{\cdot}$ for Favre-averaging. The deviations from the phase-averaged and Favre-averaged values are denoted by $\cdot'$ and $\cdot''$, respectively. Phase averaging and volume averaging are related by $\alpha\phavg{\cdot} = \vavg{\cdot}$, where $\alpha$ denotes the gas phase volume fraction. We use the symbol $\alpha_\mathrm{p}$ for the particle volume fraction. The problem under consideration is statistically homogeneous in the $y$ and $z$ directions and therefore the volume averaged equations can be expressed in one dimension. The volume averaged equations are then 
\begin{equation}
\pdt{\alpha \phrho} + \pdx{\alpha\phrho\fu_1}=0, 
\label{eq:vavgmass}
\end{equation}
\begin{equation}
\begin{split}
\pdt{\alpha \phrho\fu_1} + \pdx{\alpha\phrho\fu_1\fu_1 + \alpha \php}=&\pdx{\alpha\phavg{\sigma}_{11}}-\pdx{\alpha\phrho\favg{R}_{11}}\\ &+\frac{1}{V}\int_Sp n_1 dS - \frac{1}{V}\int_S \sigma_{1k}n_kdS, 
\end{split}
\label{eq:vavgmom}
\end{equation}
\begin{equation}
\begin{split}
&\pdt{\alpha\phrho \tilde{E}} + \pdx{\alpha\phrho\tilde{E}\fu_1+\alpha\php\fu_1}=\pdx{\alpha\phavg{\sigma_{11}}\fu_1}-\pdx{\alpha\phavg{\lambda\pdxs{T}}}\\&-\pdx{\alpha\phavg{\rho e''u''_1}} - \pdx{\alpha\phrho\tilde{R}_{11}\fu_1} + D^u + D^p + D^\mu + D^{ap} + D^{a\mu}.
\end{split}
\label{eq:vavgE}
\end{equation}
In the equations above, $\favg{R}_{11}=\wfavg{u_1''u_1''}$ is a stress due to velocity fluctuations, analogous to the classical Reynolds stress and we refer to this term as Reynolds stress throughout this paper. The continuous phase boundary is denoted by $S$, $V$ is the averaging volume, and the integrals represent the forces acting on the particle surfaces. $D^u = -1/2\pdx{\alpha\phavg{\rho u_i''u_i''u_1''}}$ is the turbulent diffusion, $D^p = -\pdx{\alpha\phavg{p'u_1'}}$ is the pressure diffusion, $ D^\mu=\pdx{\alpha\phavg{u_j'\sigma_{j1}'}}$ is the turbulent viscous diffusion, $D^{ap}=\pdx{\alpha a_1\phavg{p}}$ is the pressure-diffusion effect due to the turbulent mass flux $a=\phavg{\rho'u'}/\phrho$, and $D^{a\mu}=-\pdx{\alpha a_1\phavg{\sigma}_{11}}$ is the analogous viscous diffusion effect. An investigation of the energy balance during shock wave particle cloud interaction is outside the scope of this work, but we include the equation for completeness. The Reynolds stress appears in both \cref{eq:vavgmom,eq:vavgE}, and the terms containing it represent the forces due to velocity fluctuations and the work done by those forces, respectively. Those fluctuations can be both shear turbulence and pseudo-turbulent fluctuations. The physical processes represented by the Reynolds stress will be discussed in this paper in order to guide closure modeling.

\section{Computational method and set-up}
\label{sec:comp-set-up}

\subsection{Computational method}
The simulations in this work are performed using the compressible flow solver "CharLES", developed by Cascade Technologies. The governing equations are solved with an entropy-stable scheme on a Voronoi-mesh \citep{bres2018}, and a third order Runge-Kutta method for time stepping. A discussion of entropy stable schemes can be found in e.g. \cite{tadmor2003,chandrashekar2013}.

\subsection{Problem set-up}

\begin{figure}
	\centerline{
	\includegraphics[scale=1.0]{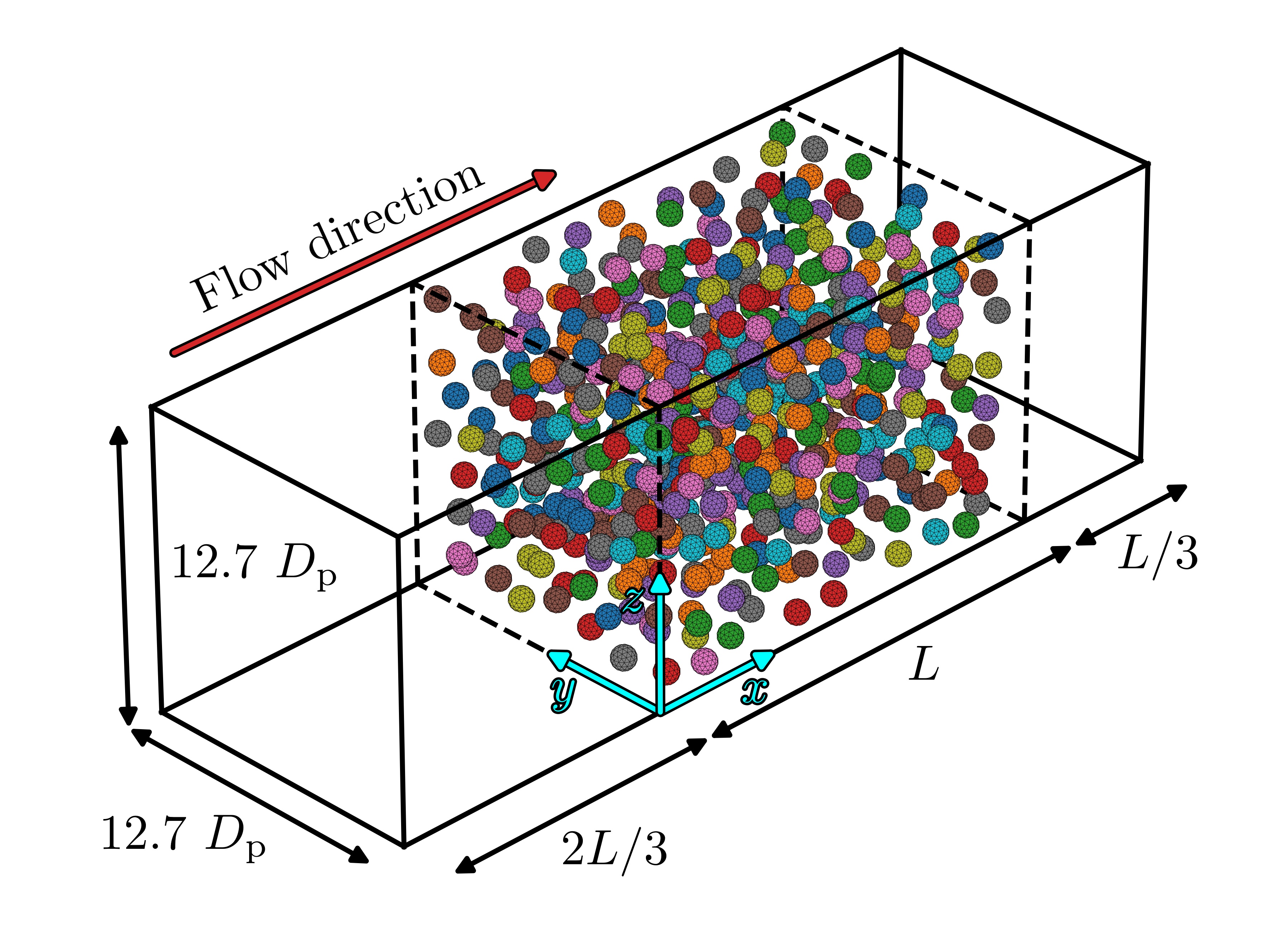}}
	\caption{Sketch of the computational domain and the particle configuration used for the simulations with the largest particles. The particles are located at $0 \leq x \leq L$, where $L=1.2\sqrt[3]{4}$ mm, and the computational domain extends $2L/3$ upstream and $L/3$ downstream of the particle cloud. The axis directions are indicated at the origin. The span-wise extent is set to a constant multiple of the particle diameter, so that $\Delta y=\Delta z = 8\sqrt[3]{4}D_\mathrm{p}$, and therefore varies depending on the particle size.}
	\label{fig:box-with-spheres}
\end{figure}

\begin{figure}
	\centerline{
	\includegraphics[]{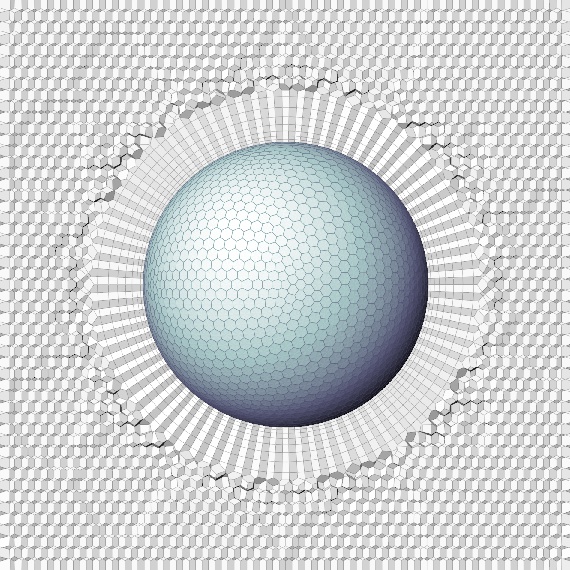}}
	\caption{Illustration of the mesh around particle, where the faces of each control-volume in this cut-plane are shaded according to the direction of their normals. There is a structured mesh around each particle extending $0.2D_\mathrm{p}$ out from the particle surface, and a Voronoi-grid in the rest of the domain. Note that the control volume sizes in this figure are adjusted for illustration purposes.}
	\label{fig:mesh}
\end{figure}

We perform numerical simulations of shock waves passing through a fixed cloud of particles, with varying shock wave Mach number, particle size and particle volume fraction. \Cref{fig:box-with-spheres} shows a sketch of the computational domain and the particle distribution. The particle cloud has length $L$, and spans the domain in the y and z directions. We denote the particle diameter by $D_\mathrm{p}$. The particle configuration in the figure is the configuration used for the simulations with the largest particle diameter. The computational grid consists of structured grids around each particle, which extend $0.2D_\mathrm{p}$ out from the particle surface, and an approximately uniform Voronoi-grid in the rest of the domain. Within this structured region, the control volume size increases geometrically with distance to the particle surface. \Cref{fig:mesh} provides an impression of the mesh around each particle. The particle positions are drawn from a uniform random distribution, so that any position within $0 \leq x/L \leq 1$ has equal probability of containing a particle. We accept a particle position if it is not closer than $1.5D_\mathrm{p}$ to any other particle center. This ensures that the structured grids do not overlap, and that there is a small distance between the structured grids where the Voronoi-grid can create a smooth transition between the two structured regions. In addition, we require that structured grids do not intersect the spanwise domain boundaries. Particles are drawn in this way until the particle volume fraction reaches the desired value. For the simulations considered in this study, the minimal number of particles in any simulation is $586$ and the maximal number is $1173$. The size of the control volumes in the Voronoi part of the mesh matches approximately the outer layer of the structured grid around each particle, and it is slightly coarsened in the regions away from the particle cloud. The total number of control volumes is roughly $6\times10^7$ for all the simulations here. On 300 cores, each simulation took roughly 24 hours to complete.

The initial condition consists of two homogeneous regions separated by a shock wave, where the pre-shock conditions are set to $\rho^0=1.2048$ kg/m$^3$, $u^0=0$ m/s and $p^0 = 1.01325\times10^5$ Pa. The post-shock conditions are determined from the shock wave strength. Post-shock quantities are used for normalization, and will be denoted with a subscript IS. For the post-shock gas velocity we omit the numeric component subscript for notational convenience. The shock wave propagates in the x-direction. The upstream boundary is set to the post-shock condition and the downstream boundary is set to a zero-gradient outlet. We apply symmetry conditions at the $y$ and $z$ boundaries. 

\Cref{tab:cases} provides an overview of the parameter combinations we simulate. The simulations will be referred to as Case I, II, ..., XII. We vary the incident shock wave Mach number between $2.2$ and $3$, particle size between $50\ \mu$m and $100\ \mu$m and particle volume fraction between $0.05$ and $0.1$. 

The analysis is conducted using the volume averaging framework. We define averaging volumes spanning the domain in the $y$ and $z$ directions, with a streamwise extent of $L/60$. These bins contain both the gas phase and the particles, and thus the particle volume fraction within a bin might deviate from the bulk particle volume fraction. The flow quantities are averaged over these bins, and we subsequently compute a moving average with a window of five bins to reduce the sensitivity of the results to the local particle configuration. 

We utilize two time-scales to compare the simulation results. These are 
\begin{equation}
\tau_L = L\left(Ma\sqrt{\gamma\frac{p^0}{\rho^0}}\right)^{-1},\quad \tau_\mathrm{p} = D_\mathrm{p}\left(Ma\sqrt{\gamma\frac{p^0}{\rho^0}}\right)^{-1},
\label{eq:timescales}
\end{equation}
where $\tau_L$ is the time it takes for the incident shock wave to travel a distance equal to the particle cloud length, $\tau_\mathrm{p}$ is the time it takes for the incident shock wave to pass over a particle and $Ma$ is the Mach number. Unless otherwise specified, the time-scale is computed using the incident shock wave Mach number in each simulation, so that the time-scales are different for the different simulations. We let $t_0$ denote the time when the shock wave is at $x=0$.

\begin{table*}
		\begin{tabular}{ccccccc}
			Case	& $Ma$ & $L_\mathrm{s}/L$ & $\alpha_\mathrm{p}$  & $Re_{\mathrm{p,IS}}$ & $n\ [\mathrm{mm}^{-3}]$ & $D_\mathrm{p}\ [\mu \mathrm{m}]$ \\ 
			\hline
			I	& $2.2$ & $0.196$ & $0.1$   & $6160$ & $191.0$ & $100$  \\ 
			II	& $2.4$ & $0.196$ & $0.1$   & $7091$ & $191.0$ & $100$  \\ 
			III	& $2.6$ & $0.196$ & $0.1$   & $7927$ & $191.0$ & $100$  \\ 
			IV	& $2.8$ & $0.196$ & $0.1$   & $8666$ & $191.0$ & $100$  \\ 
			V	& $3.0$ & $0.196$ & $0.1$   & $9309$ & $191.0$ & $100$  \\ 
			VI	& $2.6$ & $0.157$ & $0.1$   & $6292$ & $382.0$ & $79.4$ \\ 
			VII	& $2.6$ & $0.125$ & $0.1$   & $4994$ & $763.9$ & $63.0$ \\ 
			VIII& $2.6$ & $0.163$ & $0.075$ & $4537$ & $763.9$ & $57.2$ \\ 
			IX  & $2.6$ & $0.226$ & $0.05$  & $3964$ & $763.9$ & $50$   \\ 
			X	& $2.2$ & $0.099$ & $0.1$   & $3080$ & $1528$  & $50$   \\ 
			XI	& $2.6$ & $0.099$ & $0.1$   & $3964$ & $1528$  & $50$   \\  
			XII	& $3.0$ & $0.099$ & $0.1$   & $4654$ & $1528$  & $50$   \\ 
		\end{tabular} 
	\caption{\label{tab:cases} The different simulations considered in this study and key parameters. $Ma$ is the incident shock wave Mach number, $Re_\mathrm{{p,IS}}$ is the particle Reynolds number based on post incident shock values, and $n$ is the number density. $L_\mathrm{s}$ is the sight-length, as defined in \cref{sec:shockspeed}.}
\end{table*}

\section{Results}
\label{sec:results}

All simulations considered here feature the same basic flow pattern, which has been reported in a number of previous experimental and numerical studies. The most important features are the generation of the reflected shock, the generation of particle wakes, and the continuous weakening of the primary shock as it impacts on particles throughout the layer. It should be noted that the particle layer considered here is not long enough to completely dissipate the shock wave, and therefore a transmitted shock emerges from the downstream edge of the particle cloud. After the transmitted shock has moved away from the particle cloud, a flow expansion region occurs around the downstream edge of the particle cloud, where the flow transitions from subsonic ($Ma\approx 0.5-0.8$) to supersonic ($Ma\approx1.2-1.6$). The expansion region is terminated by a shock wave a short distance downstream of the particle cloud.

\subsection{Grid resolution}
\begin{figure}
	\centerline{
	\includegraphics[scale=1.0]{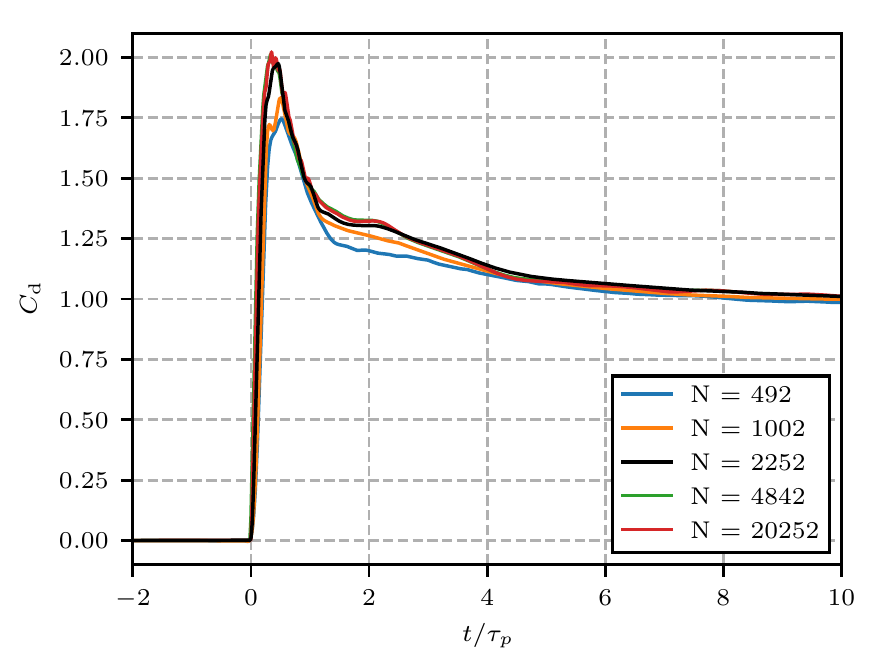}}
	\caption{Drag coefficient for an isolated particle subjected to a $Ma=2.6$ shock wave with different grid resolutions. Here, N denotes the number of points at the particle surface.}
	\label{fig:singleparticle-cd}
\end{figure}

Previous studies of shock interaction with particle arrays have estimated grid qualities by examining the drag-coefficient on a single particle \citep{mehta2016,mehta2018,hosseinzadeh2018}. It is important that the particle forces are well reproduced in the simulation, since they are central to the problem under investigation. Following the same approach, we conduct simulations of a single particle with diameter $63\ \mu\mathrm{m}$, subjected to a $Ma=2.6$ shock wave with various number of faces at the particle surface. This parameter combination is chosen because it is in the middle of the range of Mach numbers and particle sizes we have simulated. \Cref{fig:singleparticle-cd} shows the drag coefficient, as defined in \cref{eq:Cd}, as a function of time for five different grid resolutions. The drag-coefficient with $N=2252$ deviates roughly $2$\% from the highest resolution, and is a feasible resolution in terms of computational cost for the particle cloud simulations. Therefore, we apply this resolution to the simulations considered in this paper. 

\begin{figure}
	\centerline{
	\includegraphics[scale=1.0]{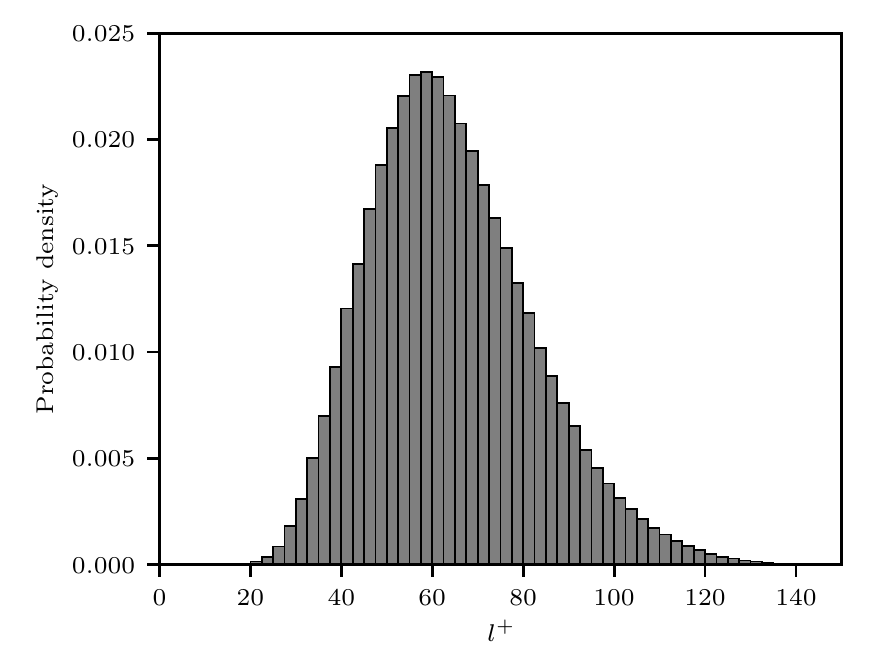}}
	\caption{Histogram of $l^+$ for case VII, with $(Ma,\ \alpha_\mathrm{p},\ D_\mathrm{p}) = (2.6,\ 0.1,\ 63\ \mu\mathrm{m})$, at $(t-t_0)/\tau_L=1.36$.}
	\label{fig:lplus-hist}
\end{figure}

In addition to the drag-coefficient, we also examine the grid-resolution in terms of the parameter
\begin{equation}
l^+ = \sqrt[3]{V_\mathrm{CV}}/l_\mathrm{viscous},
\end{equation} which is the non-dimensional grid-length scale relative to the local viscous length scale of the flow. Here $\sqrt[3]{V_\mathrm{CV}}$ is the length scale of the control volume, 
\begin{equation}
	l_\mathrm{viscous}=\sqrt{\frac{\mu}{\rho(2S_{ij}S_{ij})^{1/2}}}, 
\end{equation}
is the viscous length scale \citep{wingstedt2017} and $S_{ij}=0.5(\pdjs{u_i}+\pdis{u_j})$ is the strain rate tensor. The viscous length scale can be interpreted as the smallest length scale of the local shear flow. Thus, if the grid size is comparable to or smaller than this length scale, it is reasonable to assume that the flow is well resolved locally. The viscous length scale can be utilized independent of the state of the flow (turbulent or laminar). It should be noted that the values obtained for $l_\mathrm{viscous}$ depends on the grid-size, and $l^+$ therefore only serves as a post-simulation measure of grid-quality, as opposed to a value that can be used quantitatively to refine a mesh. \Cref{fig:lplus-hist} shows a histogram of the $l^+$ values for case VII, with $(Ma,\ \alpha_\mathrm{p},\ D_\mathrm{p}) = (2.6,\ 0.1,\ 63\ \mu\mathrm{m})$, at $(t-t_0)/\tau_L=1.36$. The middle $98\%$ of the distribution is located between $l^+=30$ and $l^+=114$. The highest values are located in the shear layer around each particle and their wakes. For problems where it is critical to resolve the turbulent energy cascade, the grid size should be comparable to $l_\mathrm{viscous}$, or if larger grid sizes are used, the smaller scales should be appropriately modeled. In this problem, it is unlikely that details of the energy cascade are very important. Therefore the requirement on $l^+$ can probably be slightly relaxed here without affecting the results considerably. Some phenomena, such as wake-wake interaction, and shock-wake interaction, might require finer resolutions than we use. However, it is not the purpose of this work to explore these phenomena in detail, and it is likely that they only have a minor effect on the results presented here.

The viscous length scales in this case are distributed between $20$ nm and $200$ nm. This means that the smallest viscous length scales are only about an order larger than typical mean free paths of air molecules. The Knudsen number based on the viscous length scale is given by 
\begin{equation}
Kn=\frac{l_\mathrm{free}}{l_\mathrm{viscous}}=\left(\frac{k_\mathrm{B}T}{\sqrt{2}\pi pD_\mathrm{p,air}}\right)\frac{1}{l_\mathrm{viscous}},
\end{equation}
where $l_\mathrm{free}$ is the mean free path of the molecules, $k_\mathrm{B}$ is the Boltzmann constant and $D_\mathrm{p,air}=3.84\times10^{-10}$ m is the effective diameter of an air molecule. Throughout most of the particle layer, the Knudsen number takes values around $0.1$. In the expansion region at the downstream end of the particle cloud, the Knudsen number increases, and around the very last particles we find values up to two. This indicates that the wakes and shear layers around particles at the downstream end of the particle cloud might be influenced by non-continuum effects, but we do not expect those effects to be very large.  

We conclude that the grid-resolution used in this study is sufficient to obtain reliable particle forces, and that it represents the local flow gradients in a satisfactory manner. 

\subsection{Shock wave attenuation}
\label{sec:shockspeed}

As the shock wave passes through the particle cloud, it is attenuated by shock reflection from the particles. The amount of attenuation depends on particle size and particle volume fraction, as well as the regularity of the particle distribution. As will be shown, the shock wave attenuation and certain mean flow trends are well characterized by a single parameter depending only on geometric properties of the particle cloud. 

\begin{figure}
	\centerline{
	\includegraphics[]{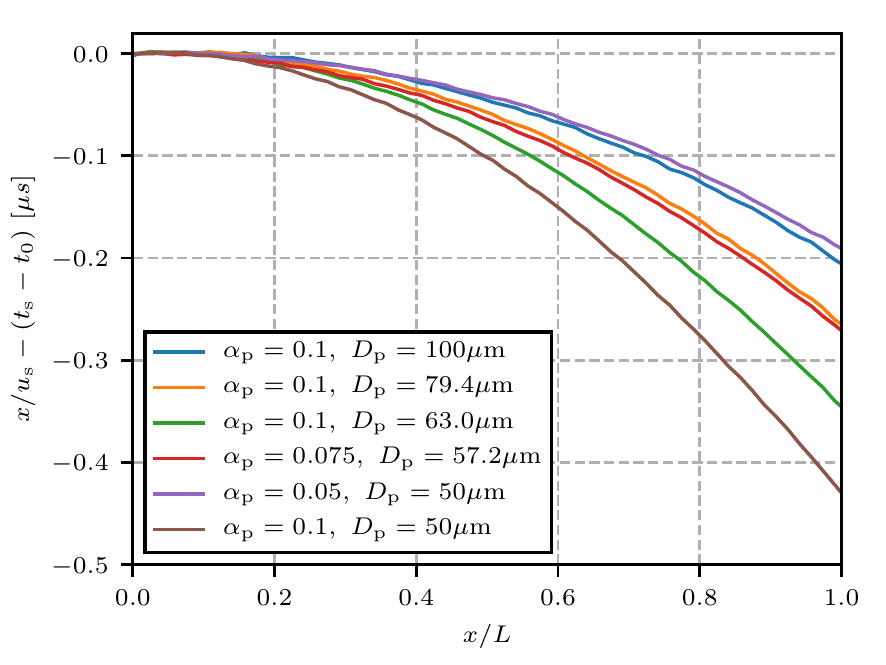}}
	\caption{Difference between undisturbed shock arrival time, $x/u_\mathrm{s}$, where $u_\mathrm{s}$ denotes the incident shock wave speed, and the obtained shock arrival time, $t_\mathrm{s}$, as a function of position within the particle cloud for an incident $Ma=2.6$ shock wave. The shock arrival time is defined as the time when the average pressure within the bin first exceeds 3 bar. }
	\label{fig:shock-arrival-time}
\end{figure}

\Cref{fig:shock-arrival-time} shows the shock arrival time as a function of distance within the particle cloud for the cases with $Ma=2.6$.  Cases VI and VIII, with ($\alpha_\mathrm{p}$, $D_\mathrm{p}$) = $(0.1,\ 79.4\ \mu\mathrm{m})$ and $(0.075,\ 57.2\ \mu\mathrm{m})$ respectively, have very similar shock speeds during the passage of the shocks through the particle clouds. Cases III and IX, with ($\alpha_\mathrm{p}$, $D_\mathrm{p}$) = $(0.1,\ 100\ \mu\mathrm{m})$ and $(0.05,\ 50\ \mu\mathrm{m})$ respectively, appear quite similar in this plot, but it can be seen that the difference between them increases with distance. The shock wave reflection imposes a strong transient force on the particles, and therefore shock wave attenuation serves as a measure of the average initial particle drag, and vice versa. The simulations do indeed show that summing up the forces on the particles during the first $\tau_\mathrm{p}$ after the shock hits each particle, yields approximately the same result for cases VI and VIII. 

\begin{figure}
	\centerline{
	\includegraphics[]{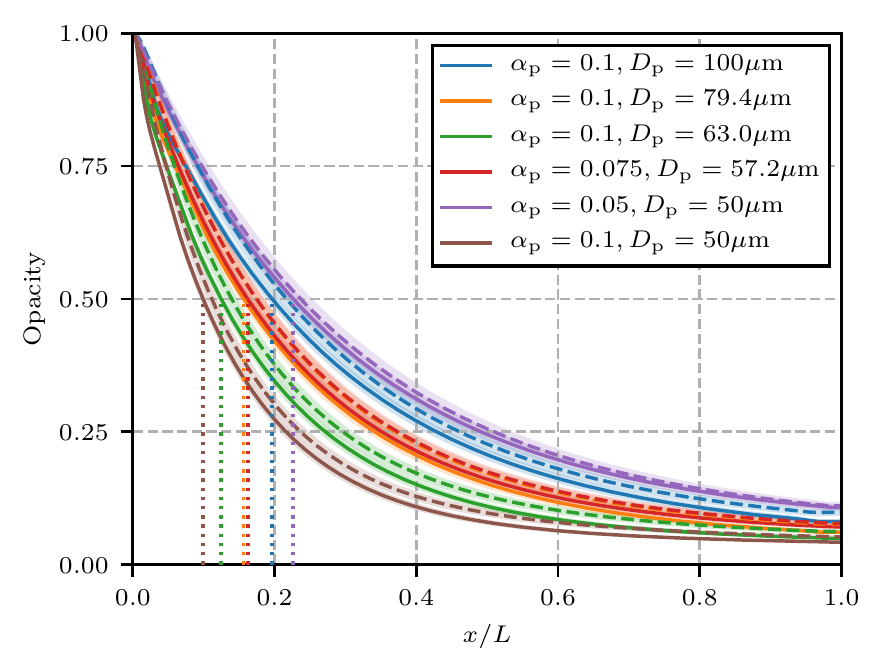}}
	\caption{Opacity for the different geometry parameters used in the simulations, as a function of distance into the particle cloud. The solid lines are the results for the particle drawing method used in this study and the dashed lines are the results for distributions without the inter-particle distance restriction. The shaded areas indicate the standard deviations for each line. The dotted lines indicate $L_\mathrm{s}$ for the different parameter combinations.}
	\label{fig:opacity}
\end{figure}

The regularity of the particle distribution can affect the shock attenuation through statistical differences in shock focusing and particle forces. For this reason, it is necessary to estimate the effect of our non-random particle distribution. We generate the particle distributions by drawing random positions satisfying two criteria that makes the particle distribution slightly more regular than a completely random distribution, as discussed in \cref{sec:comp-set-up}. To quantify this effect, we examine the area in the y-z plane that is occluded by the particles as a function of distance. The occluded area at a streamwise coordinate $x$ is the projection of all particles with streamwise coordinates less than $x$, onto a plane, accounting for overlap between the projections. We refer to the ratio of the non-occluded area to the total area as opacity, and compute it for the different geometries that we have used in the numerical simulations. \Cref{fig:opacity} shows the opacity as a function of position inside the particle cloud, where each line is the mean of 8192 realizations. The dashed lines are the corresponding opacities by only requiring that the particles are completely inside the assigned domain. The opacity for the less restrictive distribution is always slightly higher than the opacity for our particle distribution. Our particle distributions are possible realizations of the less restrictive distribution, but as there is not too much overlap between the standard deviation regions, it is clear that they are very unlikely. Since our distribution occludes more of the area over a given distance, we expect slightly stronger shock reflection and a slightly weaker transmitted shock. The effect of the restriction that the particles should not overlap the spanwise domain boundaries is examined in \cref{sec:visibilitymodel}. The result is that it increases the opacity, and is more important than the inter-particle distance. The results within this work should be interpreted with these effects in mind. Additionally, it must be emphasized that only a single realization of the particle distribution is used for the flow simulation for each parameter combination. We also note that the curves have slight bumps near $x=0$ and $x/L=1$, due to the constraints imposed on the particle distribution. We expect a similar effect for the distribution close to the spanwise domain boundaries. 

The opacities for $\alpha_\mathrm{p}=0.01,\ D_\mathrm{p}=79.4\ \mu\mathrm{m}$ and $\alpha_\mathrm{p}=0.075,\ D_\mathrm{p}=57.2\ \mu\mathrm{m}$ are very similar. These parameter combinations also resulted in a very similar shock wave attenuation, which indicates that the opacity might be used to predict some properties of shock-wave particle cloud interaction. Since the opacity curves do not seem to intersect, we use the length at which the opacity equals $0.5$ as a unique number that represents the opacity. We refer to it as the sight-length, and use the symbol $L_\mathrm{s}$. The sight-length for each configuration is given in \cref{tab:cases}, and marked in \cref{fig:opacity} for the cases with $Ma=2.6$. We see that this classification indicates that the parameter combinations $(\alpha_\mathrm{p},\ D_\mathrm{p}) = (0.1,\ 79.4\ \mu\mathrm{m})$ and $(\alpha_\mathrm{p},\ D_\mathrm{p}) = (0.075,\ 57.2\ \mu\mathrm{m})$ should be very similar, as we do observe. However, the results would indicate a larger difference between $(\alpha_\mathrm{p},\ D_\mathrm{p}) = (0.1,\ 100\ \mu\mathrm{m})$ and $(\alpha_\mathrm{p},\ D_\mathrm{p}) = (0.05,\ 50\ \mu\mathrm{m})$ than we observe. It should be noted that there is considerable standard deviation in the sight length because of the number of particles we use, so the apparent similarity between the latter two parameter combinations could be exaggerated by the specific particle distributions. A larger number of particles or an ensemble of simulations could be used to examine this in greater detail, but that is outside the scope of the current work.

The results indicate that it is possible to characterize some properties of shock wave particle cloud interaction using the sight-length. For this reason, we provide an algebraic expression that approximates this quantity in \cref{sec:visibilitymodel}.

\subsection{Mean flow}

In this section we examine the flow field during and after the shock has passed through the particle cloud. The flow quantities are phase-averaged, or Favre-averaged where appropriate, over volumes spanning the domain in the y and z directions. 

\begin{figure*}
	\centerline{
	\includegraphics[]{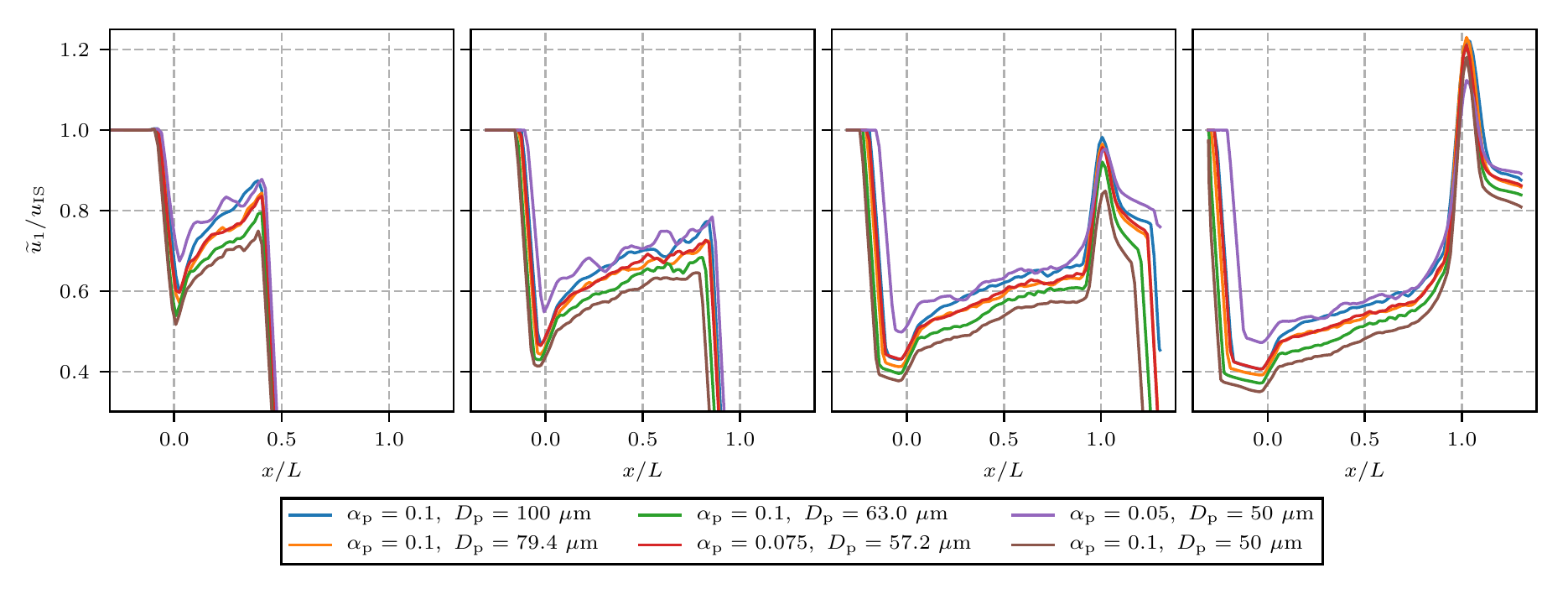}}
	\caption{Mean flow velocity with $Ma=2.6$ and different particle sizes and volume fractions at $(t-t_0)/\tau_L=0.5,\ 1.0,\ 1.5$ and $2.0$ from left to right. The particle cloud is located between $0\leq x/L \leq 1$.}
	\label{fig:M2P6-uf}
\end{figure*}

\Cref{fig:M2P6-uf} shows the normalized velocity at $(t-t_0)/\tau_L=0.5,\ 1.0,\ 1.5$ and $2.0$ for the cases with $Ma=2.6$. In the first two frames the shock wave is located inside the particle cloud. The reflected shock is visible as a sharp drop in velocity slightly before $x=0$, and the recovery shock is present around $x/L=1.1$ in the last two frames. For a given particle volume fraction, the reflected shock wave is stronger for smaller particles. This is expected based on the behavior of the primary shock discussed above, since there is a higher attenuation of the shock wave in these cases. At the upstream particle cloud edge, the mean flow speed increases rapidly over a distance equal to a few particle diameters and then has a gentler slope throughout the central region of the particle cloud. At the downstream particle cloud edge there is a strong flow expansion, and the flow speed roughly doubles over $0.9 \leq x/L \leq 1.05$ at late times. As seen in the two rightmost frames, the strength of the expansion increases with time after the shock has exited the particle layer. As noted in the discussion of grid-size above, the expansion region might be subject to non-continuum effects due to the increasing mean free path of the air molecules over the expansion region. We obtain stronger expansions for higher volume fractions and smaller particles. Thus it might be necessary to account for non-continuum effects if the volume fraction is increased or the particle diameter is decreased. The flow speed varies with volume fraction and particle sizes in the same manner as the shock speed discussed above. Again we find that the parameter combinations $(\alpha_\mathrm{p},\ D_\mathrm{p}) = (0.1,\ 79.4\ \mu\mathrm{m})$ and $(\alpha_\mathrm{p},\ D_\mathrm{p}) = (0.075,\ 57.2\ \mu\mathrm{m})$ are approximately equal, but there are slight differences between these cases in the region upstream of the particle cloud and within the expansion region. 

\begin{figure*}
	\centerline{
	\includegraphics[]{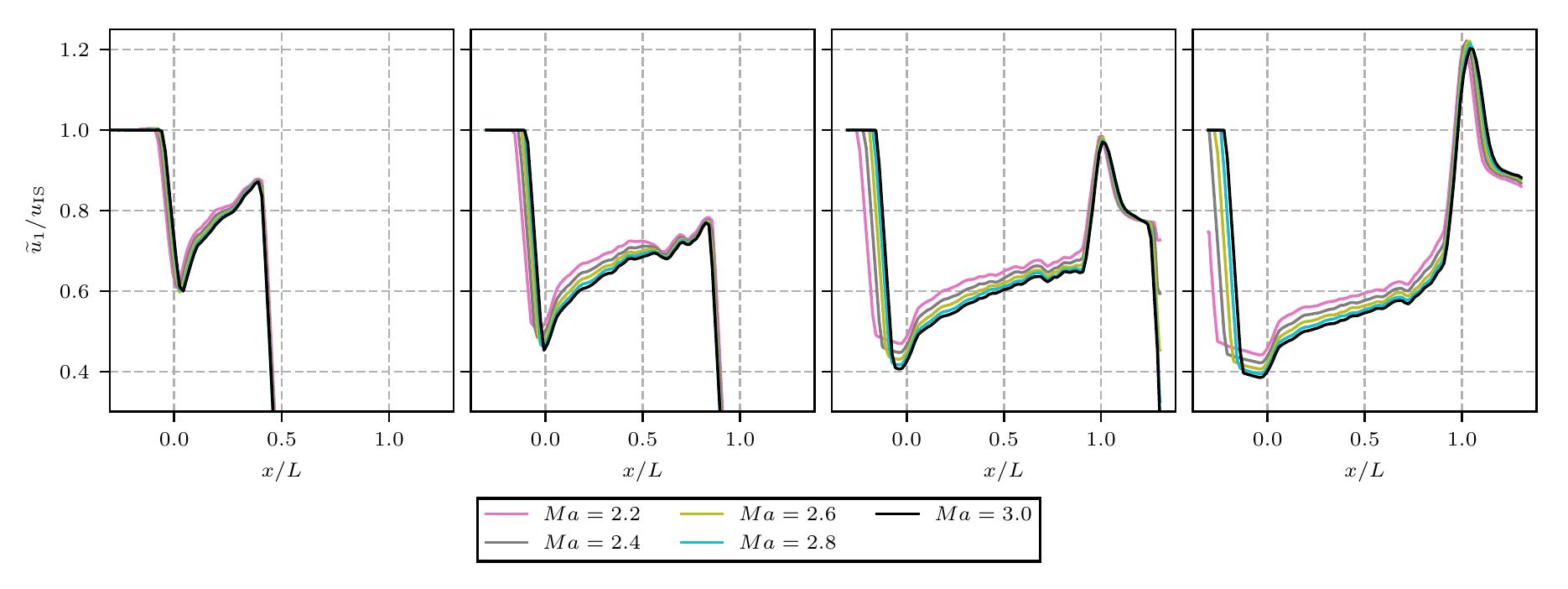}}
	\caption{Mean flow velocity with $D_\mathrm{p}=100\ \mu\mathrm{m}$, $\alpha_\mathrm{p}=0.1$ and different incident shock wave Mach numbers at $(t-t_0)/\tau_L=2$.}
	\label{fig:Machs-uf}
\end{figure*}

The variation of the mean velocity with incident shock wave Mach number is shown in \cref{fig:Machs-uf}. The normalized flow velocity within the particle layer decreases with increasing Mach number. However, the normalized velocity within a few particle diameters of the shock wave has a very weak dependence on the incident shock Mach number. The expansion region accelerates the flow to about $1.2u_\mathrm{IS}$ in all cases, and therefore the relative strength of the acceleration is larger for stronger incident shock waves. The reflected shock has a larger jump in normalized velocity with increasing Mach number. We note that the strength of the reflected shock wave increases with time over the time-frame considered here \citep{vartdal2018}.

\begin{figure}
	\centerline{
	\includegraphics[]{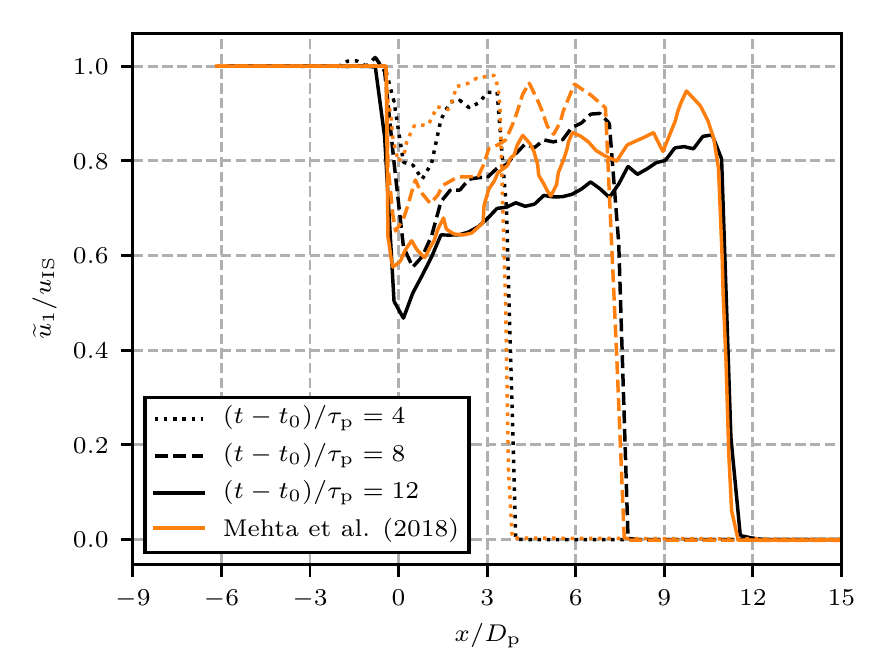}}
	\caption{Comparison of normalized velocity for case V, with $(Ma,\ \alpha_\mathrm{p},\ D_\mathrm{p}) = (3.0,\ 0.1,\ 100\ \mu\mathrm{m})$, and the inviscid simulations of \citet{mehta2018} at three times.  Black lines are the results from the simulations in this work, and orange lines are the inviscid simulation results.}
	\label{fig:mehtacomparison}
\end{figure}

It is of interest to compare the results here to those obtained in the inviscid simulation in \citet{mehta2018}. \Cref{fig:mehtacomparison} shows the results from case V, with $(Ma,\ \alpha_\mathrm{p},\ D_\mathrm{p}) = (3.0,\ 0.1,\ 100\ \mu\mathrm{m})$, at three times and the corresponding results from the inviscid simulations of \citet{mehta2018} with the same incident shock wave Mach number and particle volume fraction. On a qualitative level, the two simulations agree quite well. However, there appears to be a non-negligible difference in the reflected shock strength between the inviscid and viscous simulations, where the viscous simulations feature a stronger reflected shock wave. We emphasize again that our particle configuration is slightly more regular than that in \citet{mehta2018}, which we expect to affect the reflected shock strength. This was also observed when comparing the simulations of the inviscid face-centered cubic array in \citet{mehta2016} to the random array. We also expect that the viscous effects increase the reflected shock strength, since there will be both stronger particle drag and also a much stronger effect of Reynolds stresses. The effect of the Reynolds stress will be discussed below.

\begin{figure*}
	\centerline{
	\includegraphics[]{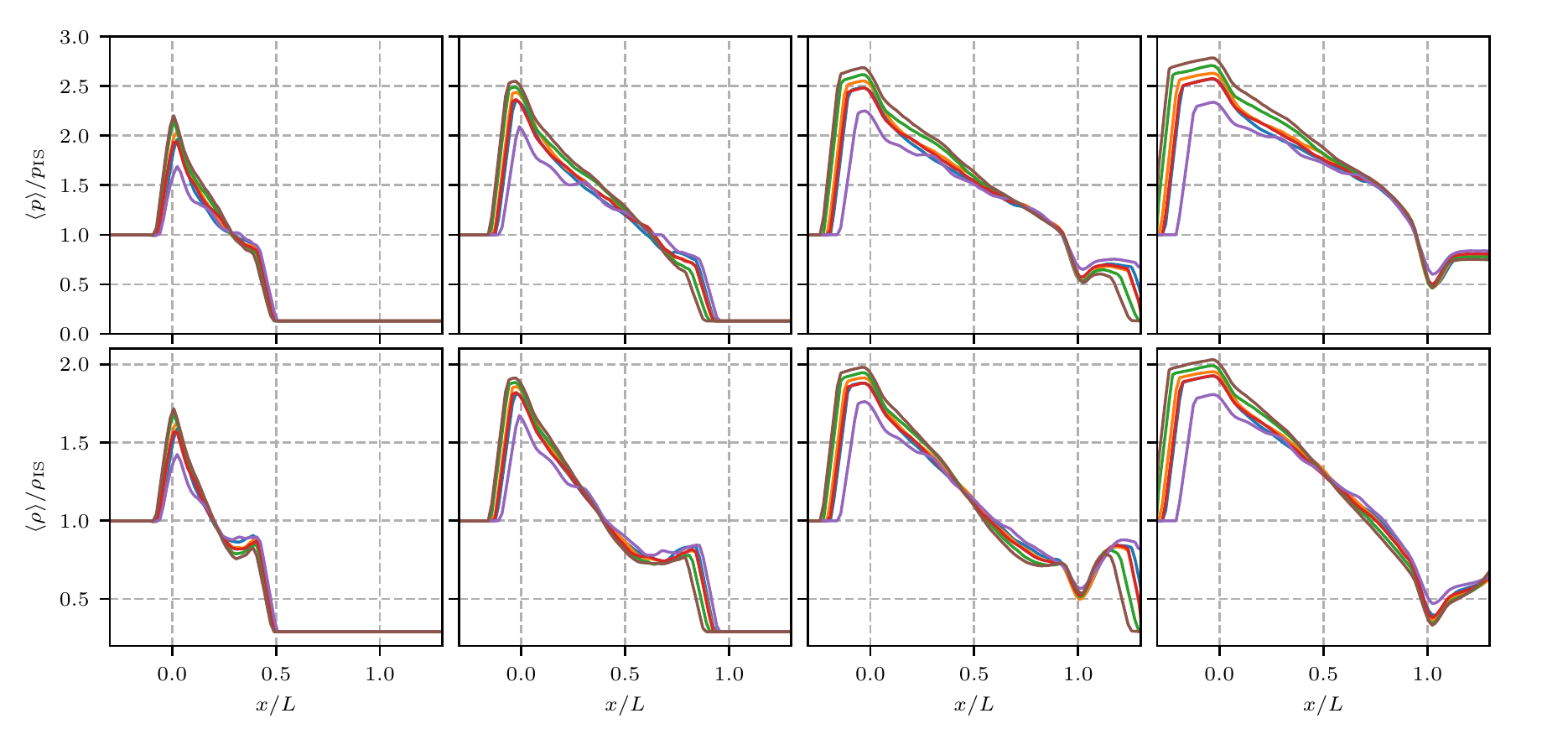}}
	\caption{Mean flow pressure (top) and density (bottom) with $Ma=2.6$ and different particle sizes and volume fractions at $(t-t_0)/\tau_L=0.5,\ 1.0,\ 1.5$ and $2.0$ from left to right. Line colors are as in \cref{fig:M2P6-uf}.}
	\label{fig:M2P6-p-rho}
\end{figure*}

The mean pressure and density profiles are shown for the different geometries in \cref{fig:M2P6-p-rho}. Both quantities display much the same behavior as the velocity field. There is a rapid change around the upstream particle cloud edge, followed by an approximately monotonic decrease throughout most of the particle cloud until quite close to the shock wave position. For the cases shown here, the pressure tends to the same level around the downstream particle cloud edge, but the pressure drop over the expansion region is significantly smaller for the lowest volume fraction case. The density profiles intersect inside the particle layer, and the configurations with higher $L_\mathrm{s}$ have lower densities at low $x$ and higher densities at higher $x$. It can be seen that the velocity, pressure and density profiles vary predictably with the sight-length. For lower $L_\mathrm{s}$ there is a higher mean velocity inside the particle layer, lower pressure and flatter density profiles. We find that case IX, with $(Ma,\ \alpha_\mathrm{p},\ D_\mathrm{p}) = (2.6,\ 0.05,\ 50\ \mu\mathrm{m})$, deviates slightly from the trend observed for the other cases. This might indicate that some of the trends we observe may be slightly different at low volume fractions, or it may be an effect of the specific particle distribution. 

\begin{figure*}
	\centerline{
	\includegraphics[]{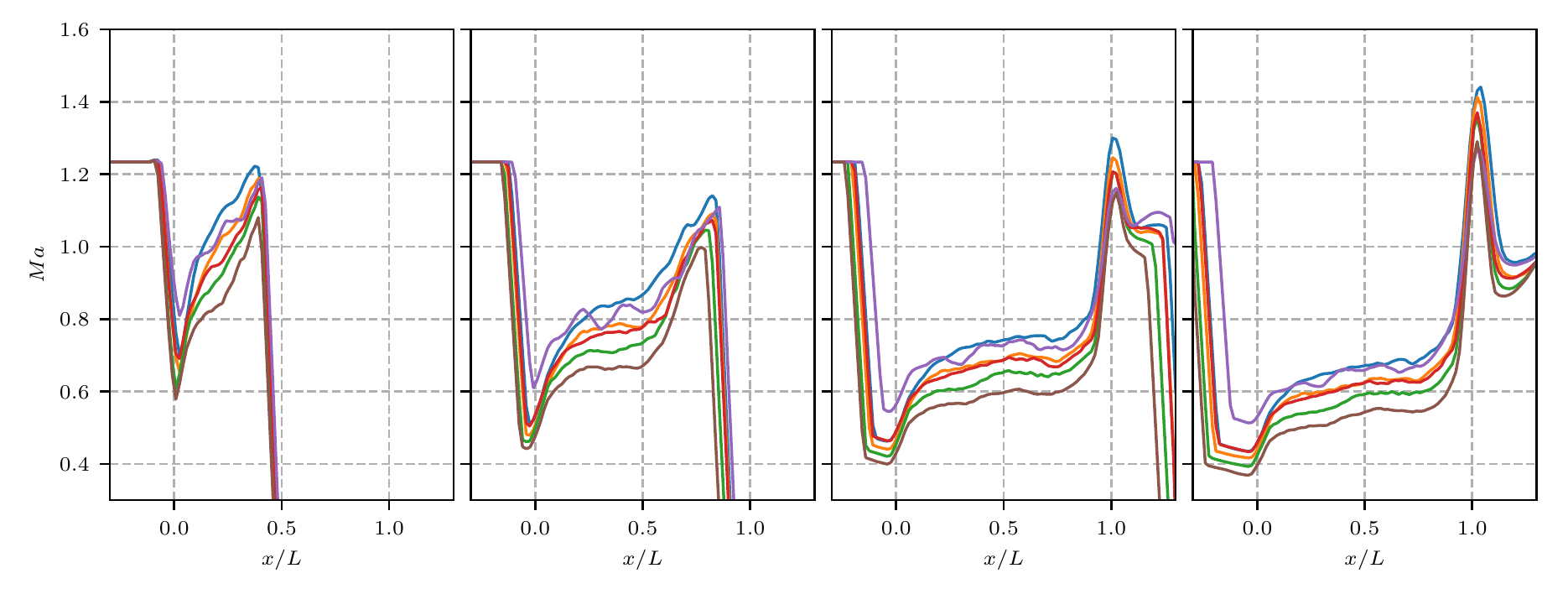}}
	\caption{Local Mach number for different geometries with incident shock wave Mach number $2.6$ at $(t-t_0)/\tau_L=0.5,\ 1.0,\ 1.5$ and $2.0$ from left to right. Line colors are as in \cref{fig:M2P6-uf}.}
	\label{fig:M2P6-Ma}
\end{figure*}

\Cref{fig:M2P6-Ma} shows the local flow Mach number for the cases with incident shock wave Mach number $2.6$. As we vary the particle volume fraction and particle diameters, we find, as expected, that the Mach number is lower for the cases where we observed a lower shock wave speed inside the particle layer. The Mach number stabilizes about $0.2L$ behind the shock wave, and has a slight positive gradient over the interior region of the particle cloud. The local Mach number drops to values around $0.5$ to $0.6$ for late times in these cases. However, it increases drastically over the expansion region, attaining values up to $1.4$ in the latest frame shown here. The transition to supersonic flow happens about one or two particle diameters upstream of the downstream cloud edge. 

\begin{figure*}
	\centerline{
	\includegraphics[]{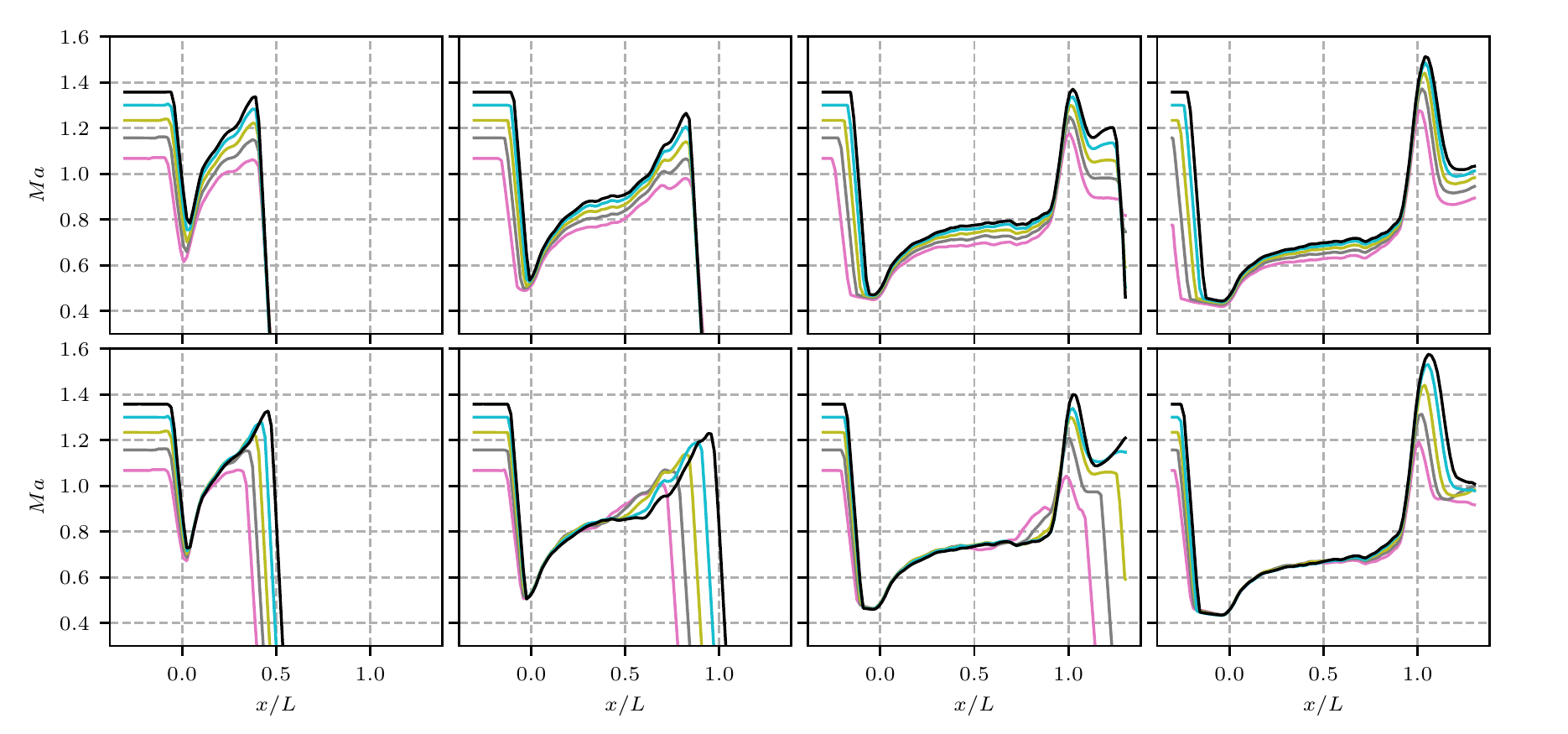}}
	\caption{Local Mach number with varying incident shock wave Mach numbers with $\alpha_\mathrm{p}=0.1$ and $D_\mathrm{p}=100 \mu\mathrm{m}$  at $(t-t_0)/\tau_L=0.5,\ 1.0,\ 1.5$ and $2.0$ from left to right (top). The bottom panels show the results at the same physical time, corresponding to using the $\tau_L$ scaling with $Ma=2.6$, i.e. the same times as shown in \cref{fig:M2P6-Ma}. Line colors are as in \cref{fig:Machs-uf}.}
	\label{fig:Machs-Ma}
\end{figure*}

\Cref{fig:Machs-Ma} shows the local flow Mach number for the cases with $\alpha_\mathrm{p}=0.1,\ D_\mathrm{p}=100\ \mu\mathrm{m}$. When compared using the time-scale based on incident shock wave speed, we find that higher incident shock wave Mach numbers result in higher local Mach numbers within the particle cloud and downstream. The expansion region is stronger for higher Mach numbers, but it appears to converge for the highest values. However, when we compare the local Mach number at the same physical time, as shown in the lower panels, we find that after the strong transient following the shock wave passage at each location, all the simulations attain the same local Mach number. We also see that the reflected shock wave has the same jump in Mach number for these simulations. In the expansion region, the results differ, and the increase in Mach number is much stronger for higher incident shock wave Mach numbers. 

\citet{regele2014} reported an average local Mach number about $0.4$ for their inviscid two-dimensional simulations with a $Ma=1.67$ shock wave and a particle volume fraction of $0.15$. Our results indicate that the local Mach number does not depend much on the incident shock wave Mach number, but has a strong dependence on particle size. The lowest average Mach number within the cloud in our cases with $Ma=2.6$ happens for $\alpha=0.1$ and $D_\mathrm{p}=50\ \mu\mathrm{m}$, where the average value at late time is $0.55$. We expect that, in addition to differences caused by the two-dimensionality, the regularity of the particle configuration in \citet{regele2014} strengthens the reflected shock wave and results in a lower local Mach number than for a random configuration. 

In summary, we find that within the central part of the particle cloud, variation of mean flow properties with particle volume fraction and particle diameter is well represented by the sight-length, which characterizes the area blockage per distance. At the downstream edge, we observe a strong flow expansion, and we find that there is a region around the upstream cloud edge that behaves differently than the central region. In these regions, which have a streamwise extent of a few particle diameters, the behavior is not predicted by the sight-length. This is because the flow field fluctuations are dynamically important in these regions, and they depend differently on volume fraction and particle diameters than the mean flow fields. This will be further discussed in the next section. 

\subsection{Velocity fluctuations}
\label{sec:velocity-fluctuations}

\begin{figure*}
	\centerline{
		\includegraphics[]{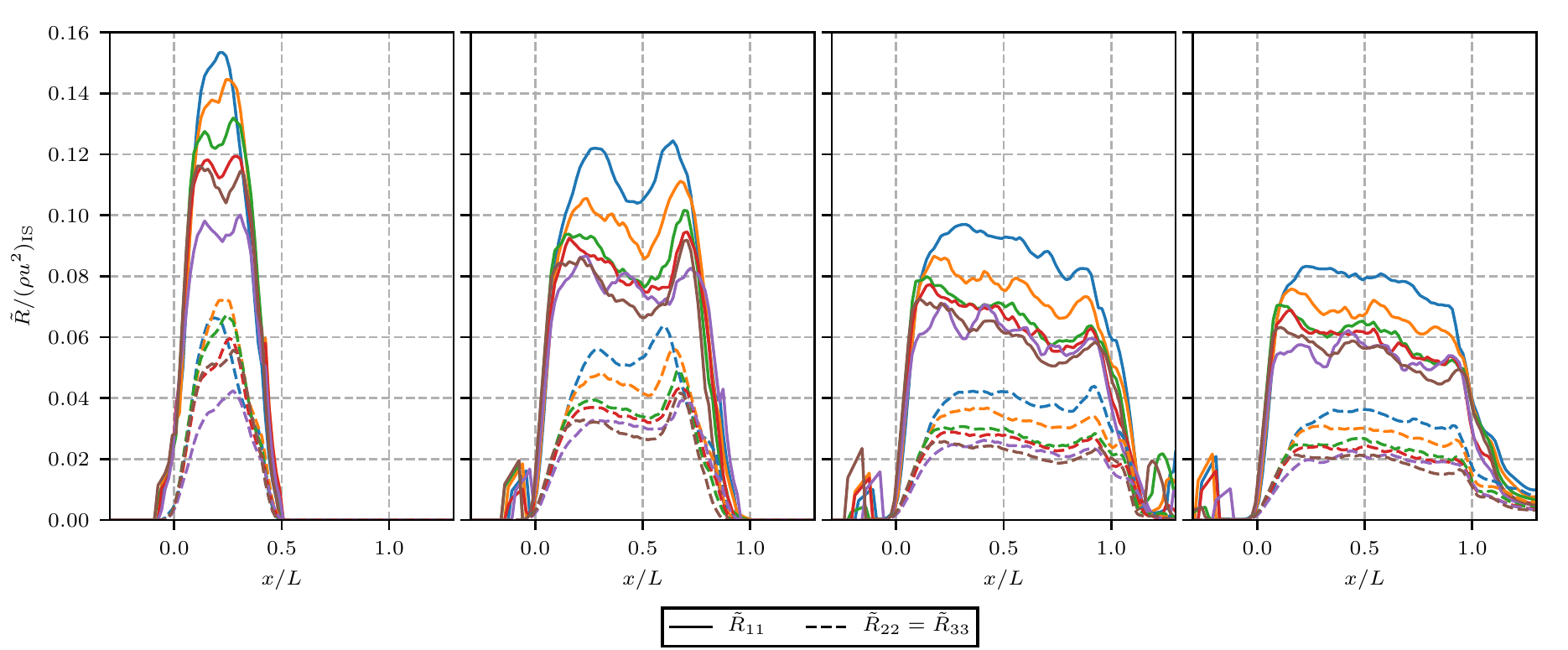}}
	\caption{Normalized streamwise Reynolds stress for different particle sizes and volume fractions at $(t-t_0)/\tau_L=0.5,\ 1.0,\ 1.5$ and $2.0$ from left to right. Line colors are as in \cref{fig:M2P6-uf}.}
	\label{fig:M2P6-R00-R11}
\end{figure*}

\begin{figure*}
	\centerline{
		\includegraphics[]{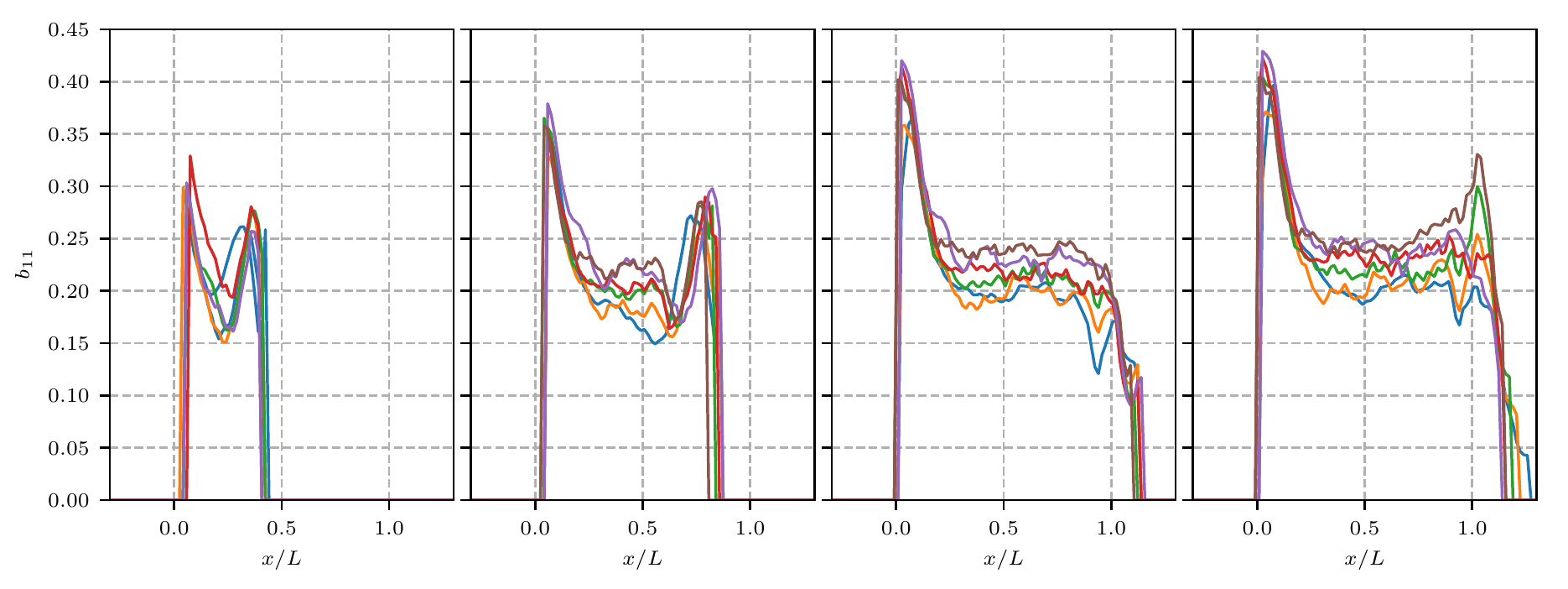}}
	\caption{Reynolds stress anisotropy for different particle sizes and volume fractions at $(t-t_0)/\tau_L=0.5,\ 1.0,\ 1.5$ and $2.0$ from left to right. Line colors are as in \cref{fig:M2P6-uf}.}
	\label{fig:M2P6-b00}
\end{figure*}

\Cref{fig:M2P6-R00-R11} shows the Reynolds stresses, $\favg{R}_{ij}$, which are the single-point, density weighted (Favre averaged), velocity fluctuation correlations, i.e.
\begin{equation}
\favg{R}_{ij}=\frac{\phavg{\rho u_i''u_j''}}{\phavg{\rho}}.
\end{equation} 
Only the streamwise component of the Reynolds stress enters the volume averaged momentum balance in this problem, but we plot the spanwise components as well since these contribute to the fluctuation kinetic energy. It can be seen that the Reynolds stress is higher at the upstream edge than in the interior of the particle cloud. It is also stronger close to the shock wave than further behind it. Upstream of the particle cloud it is zero as expected, except around the location of the reflected shock wave where the apparent Reynolds stress is an artifact of the volume averaging. The Reynolds stress is significantly higher at early times than late, and the relative magnitude does not follow the same trend as the mean fields. It has a strong dependence on the particle size, but also on the volume fraction, and it increases with both parameters. Notably, it does not vary with $L_\mathrm{s}$ in the same manner as the mean flow fields. This suggests that the characterization of the flow in terms of $L_\mathrm{s}$ only holds for a limited time, because the fluctuations should eventually affect the flow throughout the domain. In the two right-most frames, it can be seen that the Reynolds stress drops sharply over the downstream particle cloud edge. This is related to the fact that there are no longer any wakes further downstream, which are the primary contributions to the streamwise Reynolds stress in this flow. This will be discussed in more detail in \cref{sec:momentum-balance} and \cref{sec:r00model}. Similarly, the spanwise components drop because there are no particles to deflect the flow. While the Reynolds stress drops sharply, it does not vanish completely. This means that there are flow fluctuations that are advected downstream from the particle cloud.

It is worth noting the considerable magnitude of $\tilde{R}_{11}$. When scaled by twice the kinetic energy behind the incident shock wave, it reaches about $0.15$ for a particle volume fraction of $0.1$ for the early times, and decays to between $0.05$ and $0.1$ at the latest times shown here. The role of the Reynolds stress in the momentum balance is through streamwise gradients of $\tilde{R}_{11}$, and as can be seen in  \cref{fig:M2P6-R00-R11}, the gradients are very sharp around the particle cloud edges. We thus expect the Reynolds stress to play an important part of the mean flow dynamics in the regions around the upstream and downstream particle cloud edges. 

The magnitudes of the spanwise components are less than half of that of the streamwise component. We note that the off-diagonal components of the Reynolds stress in this problem are zero since the problem is constructed to have no difference between the y and z directions, and no dependence on the y and z coordinates. The velocity fluctuations are therefore statistically axisymmetric, and the coordinate axes coincide with the principal axes of the fluctuations. \Cref{fig:M2P6-b00} shows the Reynolds stress anisotropy,
\begin{equation}
b_{ij}=\favg{R}_{ij}/\favg{R}_{kk}-\delta_{ij}/3.
\label{eq:anisotropy}
\end{equation}
Since $\favg{R}_{22}=\favg{R}_{33}$, the anisotropy tensor satisfies $b_{ij}=0$ if  $i\neq j$ and $b_{22}=b_{33}=-b_{11}/2$. There is a strong anisotropy throughout the particle cloud, and it increases significantly with time. Where for the early time we observe that the maximal value of $b_{11}$ is about $0.3$, we find values as high as $0.4$ around the upstream particle cloud edge at late times, and roughly $0.225$ through most of the particle cloud. The streamwise fluctuation component increases faster with distance from the upstream edge than the other components, which ramp up over a distance of $0.2L$. We notice that the anisotropy increases towards the downstream particle cloud edge at late times, which is likely due to the strengthening of the flow expansion. This causes the difference between the velocity in the separated flow regions behind the particles in this region and the mean flow to increase, and since the spanwise fluctuations are not increased similarly, the anisotropy increases.

\begin{figure*}
	\centerline{
		\includegraphics[]{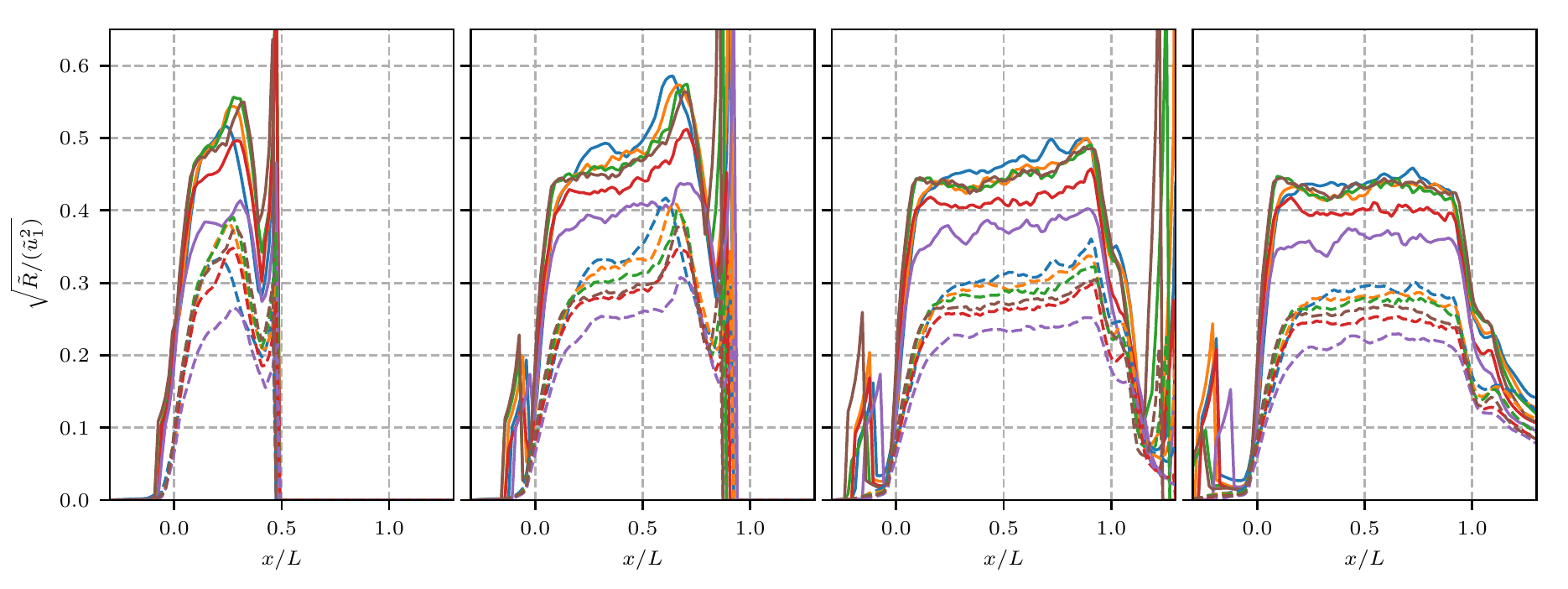}}
	\caption{Root-mean square of velocity fluctuations normalized by the mean flow density and velocity for the geometry variation at $(t-t_0)/\tau_L=0.5,\ 1.0,\ 1.5$ and $2.0$ from left to right. Line colors are as in \cref{fig:M2P6-uf}, and the dashed lines are the spanwise velocity fluctuations.}
	\label{fig:R-div-rhouu}
\end{figure*}

The importance of particle scale fluctuations in shock-wave particle-cloud interaction was previously examined in two-dimensional configurations using inviscid simulations by \citet{regele2014}, and viscous simulations by \citet{hosseinzadeh2018}, where the particle volume fraction was $0.15$. Both studies featured a $Ma=1.67$ shock wave, which is significantly weaker than the shock waves studied here. In the two-dimensional simulations, the intensity of the velocity fluctuations were comparable to the mean flow. In the inviscid simulations, the pseudo-turbulent fluctuations were approximately isotropic. Viscous effects preferentially reduced the fluctuation magnitude, with a larger reduction for the spanwise fluctuation component. An anisotropic behavior is therefore observed in both two-dimensional and three-dimensional viscous simulations, but the anisotropy is substantially stronger in the three-dimensional case.

The square root of the Reynolds stress, normalized by the mean flow velocity, is shown in \cref{fig:R-div-rhouu} for both the Mach number variation and the geometry variation with $Ma=2.6$. We note that we use bins that are longer in the streamwise direction than those used in the two-dimensional studies, which should result in larger velocity fluctuations. The normalized velocity fluctuations are insensitive to Mach number within the range we have investigated (results not shown). The streamwise fluctuation component is in addition insensitive to particle size, but the particle volume fraction has a considerable effect on the magnitude of the normalized fluctuations. By extrapolation of the trend observed with volume fraction increase, we would obtain normalized streamwise components between $0.5$ and $0.6$ for $\alpha_\mathrm{p}=0.15$ at late times, which is significantly lower than that reported in \citet{hosseinzadeh2018}. The results here differ from the previous studies which have examined fluctuations in the shock wave-particle cloud setting; we find lower fluctuation intensities than the two-dimensional studies, but significantly higher than reported for three-dimensional inviscid simulations \citep{mehta2018-b}. However, our results are quite close to that of \citet{mehrabadi2015}, who performed incompressible simulations of flow through particle suspensions up to Reynolds number 300. They reported a ratio of fluctuating to mean kinetic energy just above 0.2 for a particle volume fraction of 0.1 and a Reynolds number of 300. This ratio is slightly above 0.3 for our simulations at $\alpha_\mathrm{p}=0.1$, and we also know that it decays in time. In fact, since the local Mach numbers are not very high after about $1.5\tau_L$, it is not unreasonable to assume that the flow within the particle layer is dominated by incompressible flow phenomena after the shock-induced transients have passed.

\subsection{Upstream edge momentum balance}
\label{sec:momentum-balance}
The reflected shock is generated by the coalescence of the bow shocks from each particle, but it is sustained and strengthened by the forces acting on the flow by the particles over time. Capturing the reflected shock strength is essential for modeling shock wave particle cloud interaction since it determines the incoming flow field. If the reflected shock wave is not appropriately captured in a simulation, the mean flow fields downstream of it become incorrect, and properties such as particle drag will be computed from erroneous mean flow fields. Then only additional errors that can yield results that approximate experimental results, by introducing effects that act opposite the effect of the erroneous reflected shock wave. This situation is clearly not ideal, and we therefore stress the importance of the reflected shock wave. As will be shown, the Reynolds stress has an appreciable impact on the momentum balance around the upstream and downstream particle cloud edges. The Reynolds stress has traditionally been neglected in simulations of particle dispersion by shock waves. The results here indicate that assuming that the Reynolds stresses are negligible cannot be justified.

\begin{figure}
	\centerline{
		\includegraphics[]{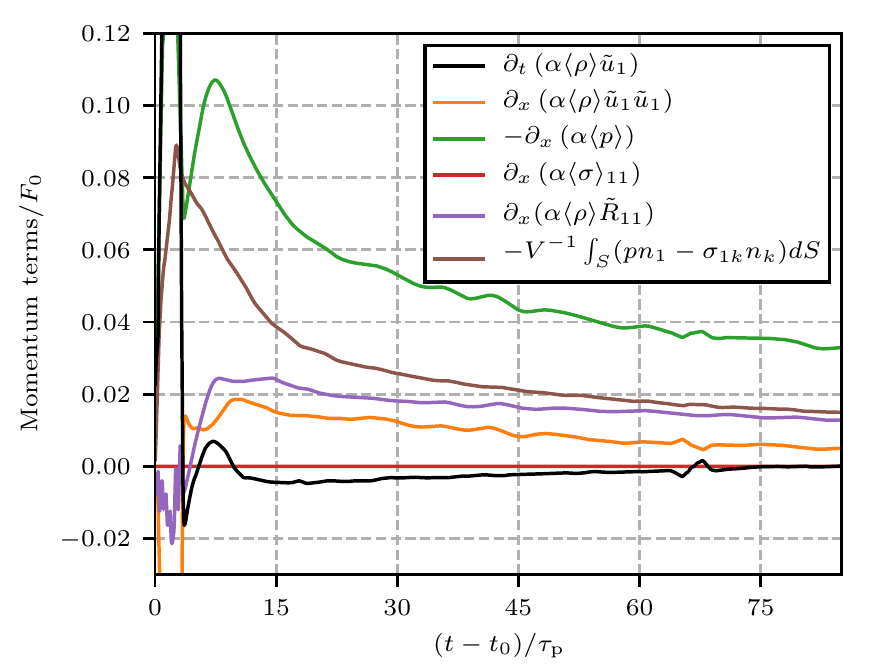}}
	\caption{Momentum balance over time at $x/D_\mathrm{p}\approx2$. For $(t-t_0)/\tau_\mathrm{p}<3.5$ the data consist mostly of artifacts from volume averaging when the shock is within the averaging volume.}
	\label{fig:momentumbalance-xddp2}
\end{figure}

We investigate the process behind the generation of the reflected shock in detail by examining the volume averaged momentum balance around the front edge of the particle cloud. \Cref{fig:momentumbalance-xddp2} shows the terms in \cref{eq:vavgmom} in a bin centered at $x/D_\mathrm{p}\approx2$ as a function of time for $(Ma,\ \alpha_\mathrm{p},\ D_\mathrm{p}) = (2.6,\ 0.1,\ 63\ \mu\mathrm{m})$. The bin size is $L/60$, but the values are averaged over five bins so the values presented here are influenced by a volume spanning about $2.5D_\mathrm{p}$ in the streamwise direction. The terms are normalized by 
\begin{equation}
F_0=(\rho u)_\mathrm{IS}/\tau_\mathrm{p}.
\end{equation}
We choose to show the results for this particular parameter combination because it is in the middle of the range of incident shock wave Mach numbers and particle sizes we have simulated, and the observations made for this case are representative of the other cases. Until $(t-t_0)/\tau_p\approx3.5$, the shock is within the region influencing this bin, and during this time, the results are dominated by artifacts from the averaging method. The most important processes occurring during this time are the shock wave reflections by the particles within the bin. Since the reflected shock waves do not have time to interact with each other, except where the particles are very close to each other, the resulting flow field is essentially just a superposition of individual shock wave-sphere interactions. After the shock has passed out of the bin, we find that the momentum-balance can be split into two different time-intervals. In the first interval there is a transient in the strength and relative importance of the different terms in the volume averaged momentum equation. The second has a slow decay of the strength of all the terms over time. It can be seen that the largest terms are the pressure gradient and the drag on the particles. The Reynolds stress contribution is initially quite a bit lower than the particle drag and pressure gradient, but it decays less over time than the other terms. After the strong initial transient it becomes comparable to the particle drag.

\begin{figure*} 
	\centerline{
		\includegraphics[]{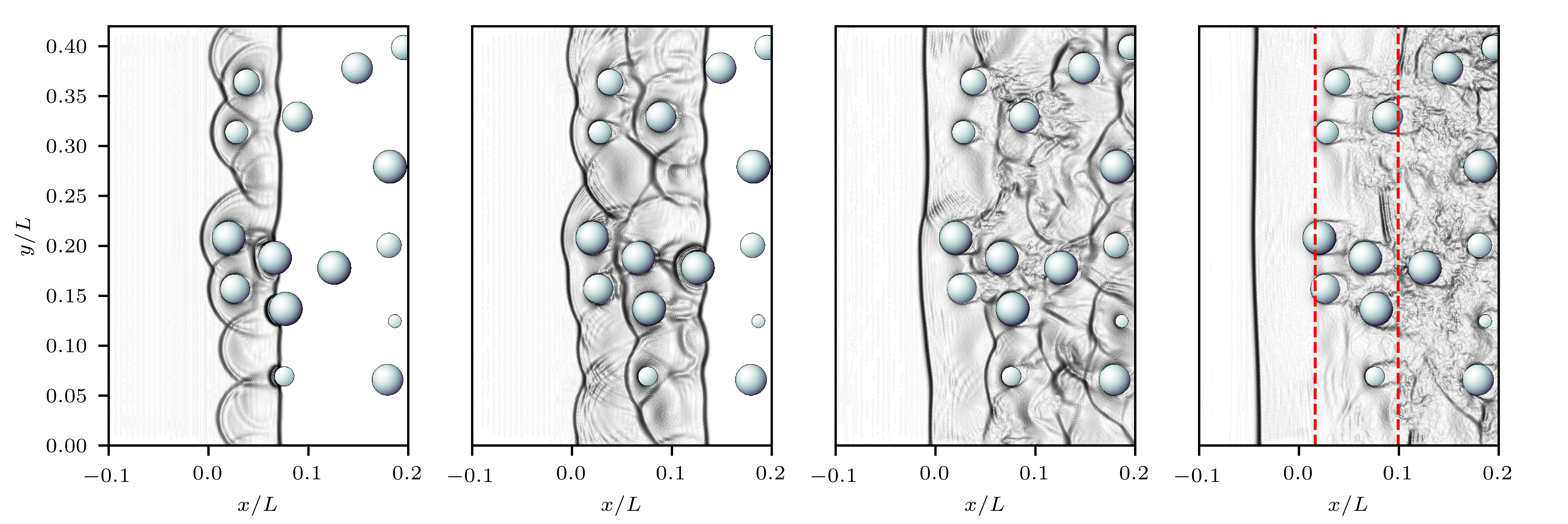}}
	\caption{Numerical schlieren-images on a slice through the middle of the domain for the case with $(Ma,\ \alpha_\mathrm{p},\ D_\mathrm{p})=(2.6,\ 0.1,\ 63\ \mu\mathrm{m})$ at times $(t-t_0)/\tau_p=2$, $4$, $8$, and $16$, from left to right. In the fourth image, the red dashed lines show the size and location of one of the averaging volumes used in the analysis of the flow field.}
	\label{fig:schlieren-four-early-times}
\end{figure*}

An impression of the processes occurring during the first transient can be obtained using numerical schlieren images, as shown in \Cref{fig:schlieren-four-early-times}. These images are of an xy-plane through the middle of the domain at $(t-t_0)/\tau_\mathrm{p}=2,\ 4,\ 8$ and $16$. In the fourth image, the size and location of the averaging region used in \cref{fig:momentumbalance-xddp2} are indicated by the red dashed lines. The first time-interval, in which the strength of the momentum balance terms change quickly, is similar to the time-period spanned by these plots. During this time, there are numerous reflected shocks within the averaging volume, but they are almost completely gone at $(t-t_0)/\tau_\mathrm{p}=16$. As the reflected shocks from the particles propagate upstream, the pressure rapidly builds up upstream of the bin. Subsequently, the pressure within the bin builds up due to the bow-shocks from particles further downstream. This process causes the rapid change in pressure gradient around the minimum at $(t-t_0)/\tau_\mathrm{p}\approx 8$. Over the same time, the mean flow velocity decreases, and therefore the magnitudes of the particle forces are reduced.

\begin{figure}
	\centerline{
		\includegraphics[]{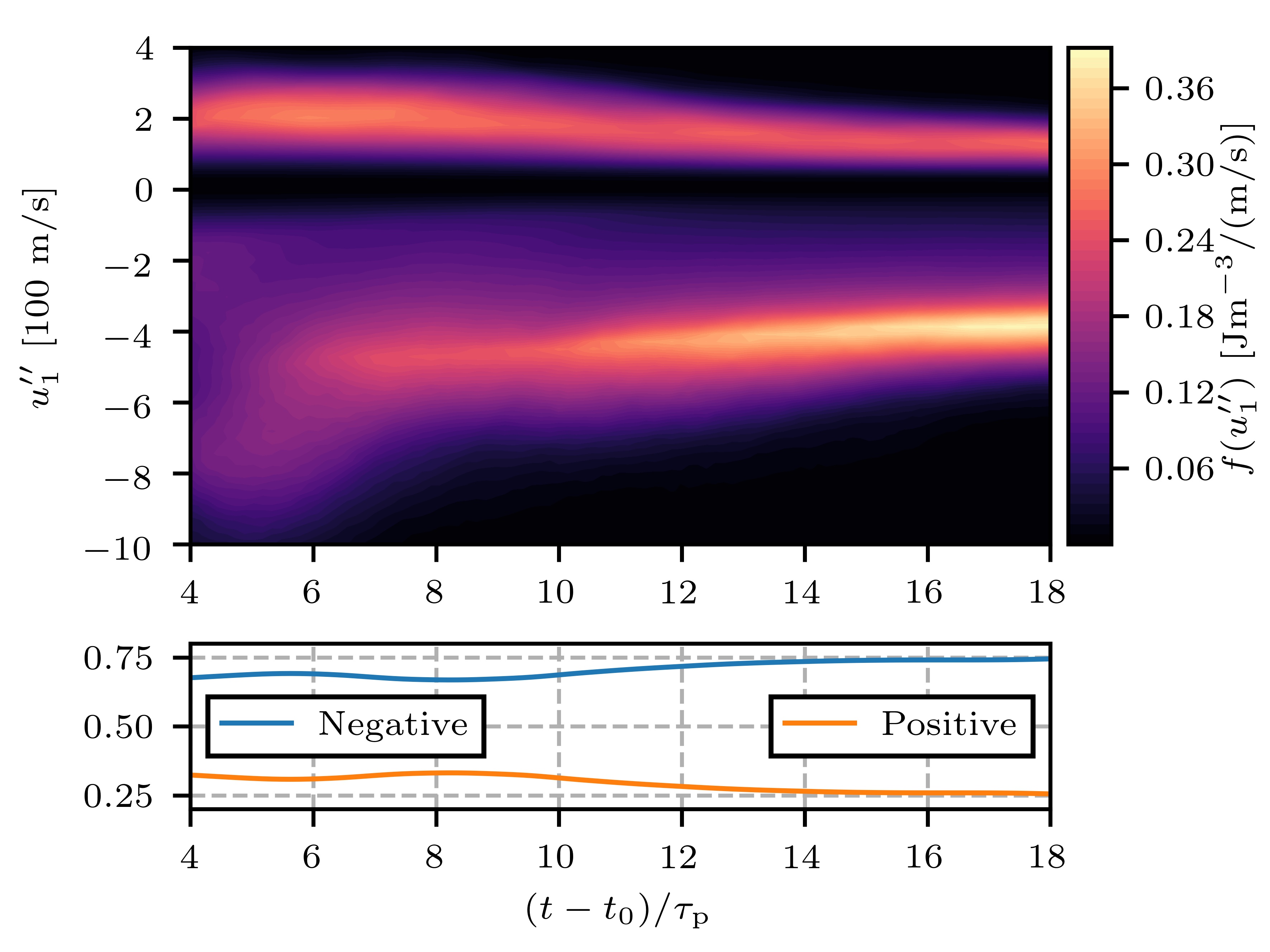}}
	\caption{Contour plot of the function $f$, as defined in \cref{eq:R00integral}, as a function of velocity fluctuation and time. The lower panel shows the total contribution to $\tilde{R}_{11}$ by positive and negative velocity fluctuations.}
	\label{fig:uupdfu}
\end{figure}

The build-up of particle wakes also occurs during this time period, as can be seen in \cref{fig:schlieren-four-early-times}. The particle wakes and the shock reflections are the main contributions to the velocity fluctuations within each bin, and thus make up the Reynolds stress in the volume averaged equations. Insight into which processes that contribute to the Reynolds stress can be obtained by consideration of the function $f(u''_1)$, defined as
\begin{equation}
\phavg{\rho}\tilde{R}_{11} = \frac{1}{V_\mathrm{gas}}\sum_{i=1}^{N_\mathrm{CV}} \left(\rho u''_1 u''_1\right)^iV_{CV}^i \approx \int_{-\infty}^{\infty}f(u''_1)du''_1.
\label{eq:R00integral}
\end{equation}
Here $V_\mathrm{gas}$ is the volume of the gas within the bin, $N_\mathrm{CV}$ is the number of control volumes in each bin, and the superscript $i$ denotes control volume number $i$. This function is the contribution to the streamwise Reynolds stress by streamwise fluctuations of a certain magnitude. Contours of $f$ as a function of $u_1''$ and time are shown in \cref{fig:uupdfu}. The figure also shows the total contribution to the integral by positive and negative fluctuations. Until $(t-t_0)/\tau_\mathrm{p}\approx 8$ there is a comparable contribution to $\tilde{R}_{11}$ by negative fluctuations with magnitudes from $100$ m/s to $1000$ m/s, due to shock wave reflection from the particles. There is also a narrow band of positive fluctuations making up about $30\%$ of $\tilde{R}_{11}$. Later, the contribution by negative fluctuations are primarily by fluctuations between $-200$ m/s and $-600$ m/s. The very high fluctuation magnitudes can only be caused by the particle wakes, since at late times there are no longer any shocks within the particle layer. The lower panel shows that more than $70$\% of $\favg{R}_{11}$ can be attributed to the negative fluctuations, and the percentage increases slowly with time. The contribution to $\tilde{R}_{11}$ from positive fluctuations are predominantly due to two effects. The first of these effects is the local flow acceleration around particles in regions where the local particle distribution is denser than average. The second effect is caused by the separation regions behind each particle, which when added together amount to quite large regions with very low flow speeds. These regions shift the average velocity away from the "free" flow velocity. As a consequence, there are very few regions in the flow which have velocities equal to the mean velocity, and this results in contributions to $\tilde{R}_{11}$ from regions of smooth "free" flow. This has consequences for modeling since the mean slip velocity, if assumed to be equal to the volume-averaged velocity, is lower than it should be. There is a similarity between this problem and the self-induced flow disturbance problems examined in \cite{horwitz2016,balachandar2019}, in the sense that flow disturbance induced by the particle, i.e. the separated flow, affects how particle drag should be calculated. In \cref{sec:r00model}, we propose a simple model for the Reynolds stress that also provides a ratio of the average flow speed to the free flow speed. The free flow speed is higher than the average flow speed, and it may be more suitable for calculating particle drag.   
 
It is clear that the momentum balance around the upstream particle cloud edge has strong contributions from both particle drag and Reynolds stress. Together these balance about two-thirds of the pressure gradient. The (time-dependent) strength of the reflected shock wave thus depends strongly on the particle forces and the velocity fluctuations caused by shock wave reflection and flow separation. In Eulerian-Lagrangian or Eulerian-Eulerian methods, these quantities require careful modeling so that the reflected shock wave, and therefore the incoming flow, is correct. The viscous simulations show fluctuations with much higher magnitudes than inviscid simulations because of flow separation. Therefore viscosity has an appreciable effect on the reflected shock strength.

\subsection{Particle drag}
\label{sec:particledrag}
The forces acting on the particles are one of the most important aspects in modeling of dispersed flows when the particle relaxation time is large compared to the mean flow time scale. The initial force history, and the distribution of peak drag coefficients, in shock wave particle cloud interaction has been characterized in \citet{mehta2018}. \citet{theofanous2018} performed particle-resolved inviscid simulations with an immersed boundary method, and found a dispersive behavior at the downstream end of the particle cloud, in agreement with experimental results. They also pointed out that the opposite behavior is typically seen in Eulerian-Lagrangian or Eulerian-Eulerian simulations of shock-wave particle interaction, i.e. a particle accumulation at the downstream edge. Those previous studies both utilized inviscid simulations. It is clear from the discussion above that viscous effects affect the flow field. The particle wakes in particular differ in viscous and inviscid simulations, and therefore the particle forces differ after the particle wake has developed. For this reason, we examine the particle drag during the time where the particle wakes have developed. 

\begin{figure}
	\centerline{
		\includegraphics[]{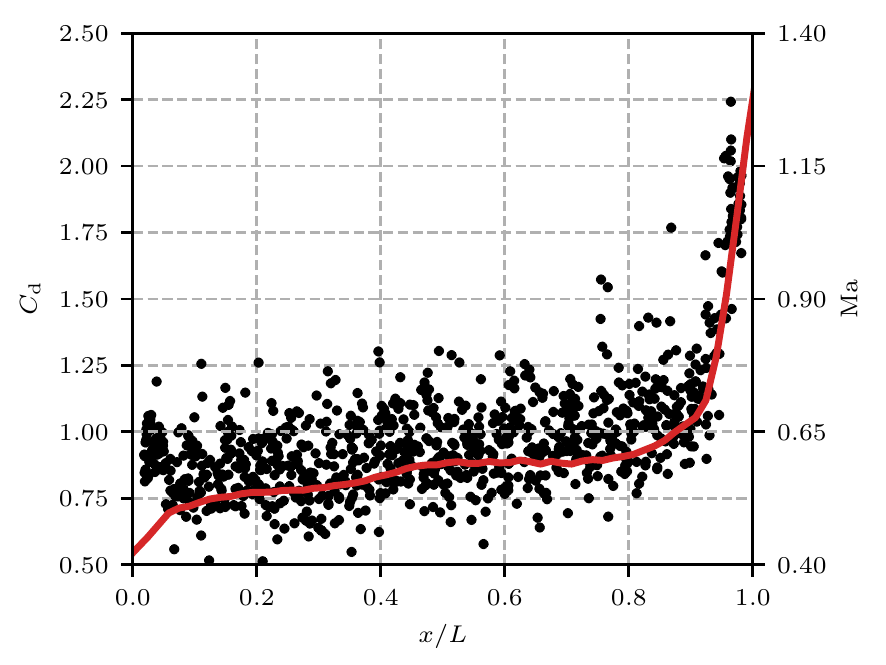}}
	\caption{Black disks: time-averaged drag-coefficients over the time interval after $0.5\tau_L$ after the shock wave has passed each particle in the case with $(Ma,\ \alpha_\mathrm{p},\ D_\mathrm{p})=(2.6,\ 0.1,\ 63\ \mu\mathrm{m})$ as a function of position. Solid red line: mean flow Mach number at $(t-t_0)/\tau_L = 2$.}
	\label{fig:cd-vs-x-caseVII}
\end{figure}
The particle drag coefficient is defined as
\begin{equation}
C_d = \frac{\int_{S_i} (p+\sigma_{1k})n_kdS_i}{0.5\phavg{\rho}\favg{u}_1^2A_\mathrm{p}},
\label{eq:Cd}
\end{equation}
where $A_\mathrm{p}$ is the projected area of the particle in the direction of the flow, and $S_i$ denotes the surface of the particle. The time-averaged drag-coefficient, averaged in time starting at $0.5\tau_L$ after the shock wave has passed, is shown in \cref{fig:cd-vs-x-caseVII} for $(Ma,\ \alpha_\mathrm{p},\ D_\mathrm{p})=(2.6,\ 0.1,\ 63\ \mu\mathrm{m})$ as a function of position. The drag coefficient increases slowly with distance throughout the particle cloud, until about $x/L=0.9$, where the drag coefficient abruptly increases and reaches a maximum of $2.25$. While the average drag coefficient increases with distance, the averaged particle forces decrease with distance, because the kinetic energy of the flow decreases with distance. Similarly, \cite{mehta2018} found that the peak particle forces decrease with distance within the particle cloud. So for most of the particle layer, the peak acceleration and the averaged forces decrease with distance. There is a slight difference between the very first particles and those a bit further in, because the first particles are exposed to the smooth incident flow rather than the chaotic flow further inside the particle cloud. The abrupt increase in drag-coefficient around the downstream edge can be explained by looking at the mean flow Mach number, which is shown as a red line in \cref{fig:cd-vs-x-caseVII}. The increase in drag-coefficient coincides with an increase in local Mach number as the flow expands and becomes supersonic. The abrupt increase in drag in the transonic region is consistent with previous findings that have examined single-particle drag as a function of Mach number \citep{bailey1971,nagata2016}. We obtain significantly higher drag coefficients than reported in those single-particle studies. The wide distribution of drag coefficients is caused by the random particle distribution, which creates local flow acceleration and deceleration. However, only the minimal drag coefficients we obtain are close to those reported in the mentioned studies, so there is a clear difference between isolated particle drag and the drag within a particle cloud. However, in \cref{sec:r00model} we propose a correction to the particle drag. Using this correction, the particle drag coefficients are reduced, which brings them closer to the single particle drag drag-coefficients. 

\begin{figure*}
	\centerline{
		\includegraphics[]{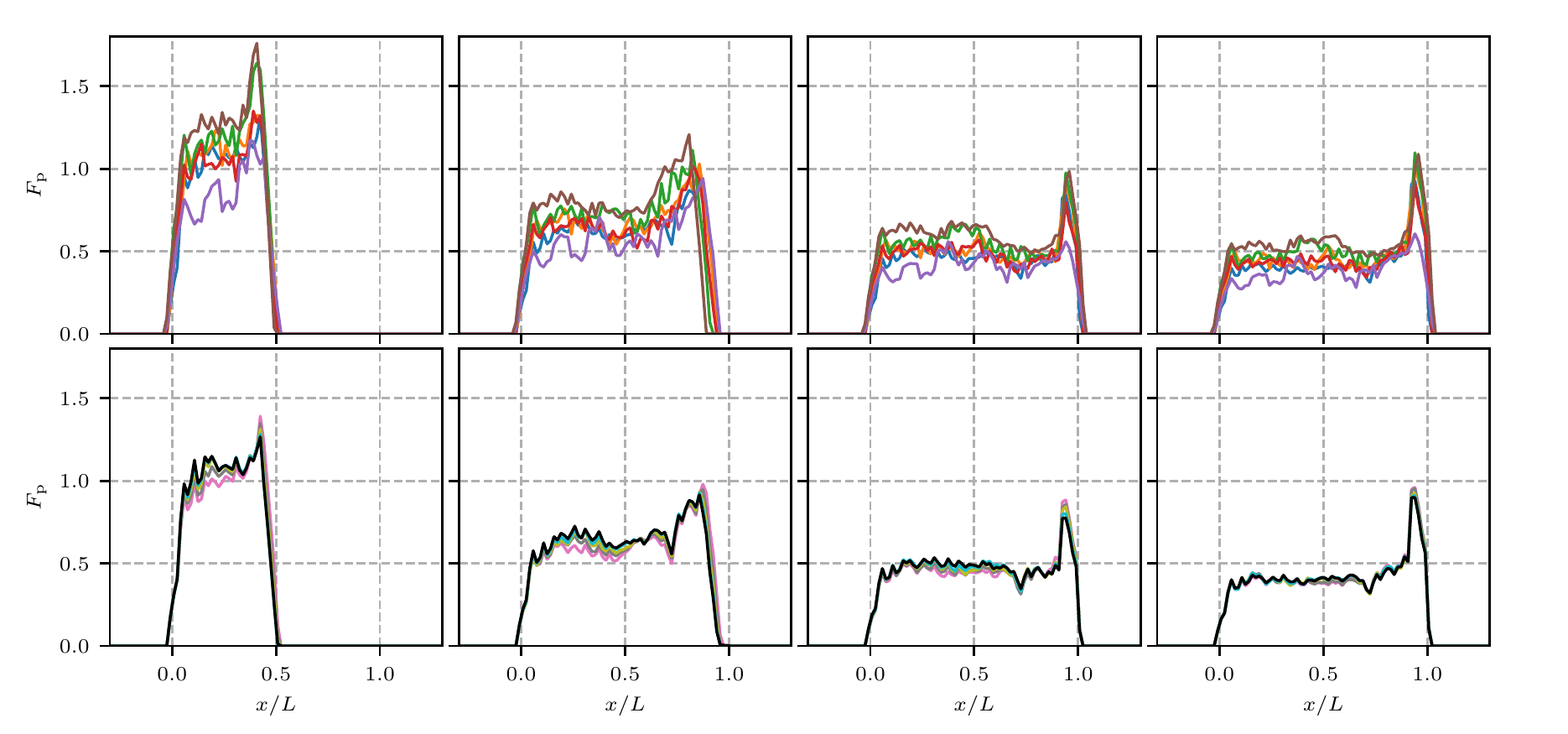}}
	\caption{Total force exerted on the particles per unit volume at $(t-t_0)/\tau_L=0.5,\ 1.0,\ 1.5$ and $2.0$, for $Ma=2.6$ and the different geometries (top), and different Mach numbers for $\alpha_\mathrm{p}=0.1$, $D_\mathrm{p}=100\ \mu$m (bottom). Line colors are as in \cref{fig:M2P6-uf} and \cref{fig:Machs-uf}.}
	\label{fig:ftx}
\end{figure*}

The total streamwise forces exerted on the particles per unit volume, normalized by post-shock momentum per $\tau_L$, i.e. 
\begin{equation}
F_\mathrm{p}= \frac{\tau_\mathrm{L}}{V\left(\rho u\right)_\mathrm{IS}}\int_S pn_1 + \sigma_{1k}n_kdS,
\end{equation}
as a function of distance at $(t-t_0)/\tau_L=0.5,\ 1.0,\ 1.5$ and $2.0$, is shown in \cref{fig:ftx}. The particle force per unit volume increases with increasing particle volume fraction and with decreasing particle diameter. The variation with particle volume fraction is stronger than with particle size within the range we have simulated. The particle force is significantly higher at early times than late. Furthermore, it increases drastically at the downstream cloud edge at late times. The exception to this trend is the case with $(Ma,\ \alpha_\mathrm{p},\ D_\mathrm{p})=(2.6,\ 0.05,\ 50\ \mu\mathrm{m})$. In that case, which has the lowest particle volume fraction, and the smallest particles, the increase at the downstream edge is quite moderate. The variation with incident shock wave Mach number is shown in the lower panels of \cref{fig:ftx}. As the Mach number is increased, the particle forces increase, and when normalized by post-shock momentum and a time-scale based on the shock speed, the forces are very similar, especially at the later times. 

\section{Modeling}
\label{sec:modeling}
In this section, we propose two algebraic models based on the observations made in the preceding sections. The first model is for the sight-length, and can be used to estimate combinations of $\alpha_\mathrm{p}$ and $D_\mathrm{p}$ that behave similarly. We also use this model to quantify the effect of the regularity of the particle distribution used in the flow simulations. The second model is an algebraic, single-parameter expression for the streamwise Reynolds stress. It reliably predicts the streamwise Reynolds stress intensity over time for all the simulations in this work. Accurate Reynolds stress models are important in Eulerian-Lagrangian and Eulerian-Eulerian models for dispersed flow, due to the importance of the fluctuations in the flow dynamics.

\subsection{Sight-length} 
\label{sec:visibilitymodel}
\begin{table}
		\begin{tabular}{c c c}
			Restrictions & $A$ & $B$\\
			\hline
			Inter-particle distance and spanwise boundaries & $-1.012$ & $0.4775$\\
			Inter-particle distance & $-0.5191$ & $0.4098$\\
			Spanwise boundaries & $-0.5313$ & $0.4703$\\
			None & $-0.3317$ & $0.3957$\\
		\end{tabular}
		\caption{Best fit for the constants $A$ and $B$ in \cref{eq:sight-length} with and without restrictions on inter-particle distance and spanwise domain boundaries.} 
		\label{tab:AB}
\end{table}

\begin{figure}
	\centerline{
	\includegraphics[]{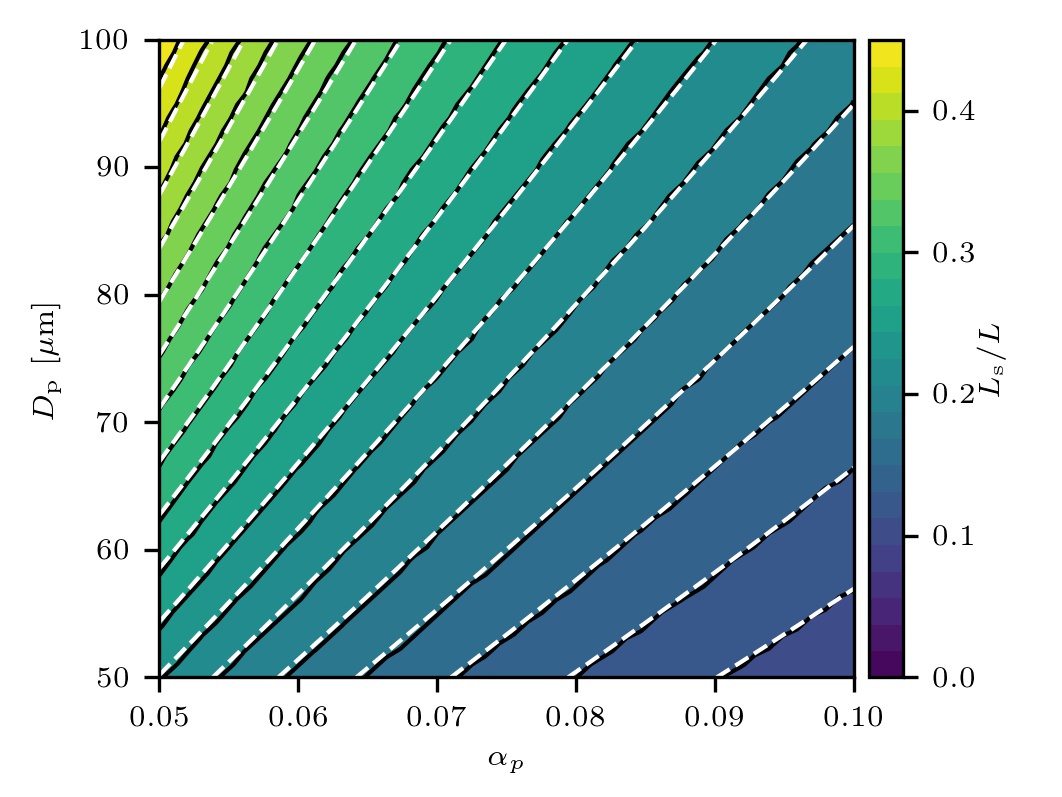}}
	\caption{Sight-length into the particle cloud for various particle sizes and particle volume fractions. The black lines are data contour lines, while the white dashed lines are the corresponding contours of \cref{eq:sight-length}.}
	\label{fig:sight-length-contours}
\end{figure}

Based on the results presented in \cref{sec:shockspeed}, we found that the shock wave attenuation throughout the particle cloud could be well characterized by the sight-length. Simulations with similar sight-lengths had similar shock wave attenuation over a given distance. Many of the mean flow quantities were also found to be predictable based on the sight-length. We thus seek an expression for the sight length in terms of the particle volume fraction and the particle diameter. To this end, we compute the sight-length for a range of particle volume fractions between $0.05$ and $0.1$ and particle sizes between $50\ \mu$m and $100\ \mu$m, by sampling the particle distribution 10240 times for each parameter combination. We fit a model on the form  
\begin{equation}
L_\mathrm{s} = \left(A+B/\alpha_\mathrm{p}\right)D_\mathrm{p},
\label{eq:sight-length}
\end{equation}
where $A$ and $B$ are the model constants to be fitted to the data. We do this using a non-linear least squares method. This model form goes to infinity as the particle volume fraction goes to zero, and to 0 for a particle volume fraction of $\alpha_\mathrm{p}=\left|B/A\right|$. The model is however not intended for volume fractions outside the range  studied in this work, and so the latter limit is of little consequence. To quantify the effect of the additional constraints we have put on the particle distribution, we also sample the particle distributions in a much larger spanwise domain and compute the sight-length in a subset of this larger domain. We refer to this setting as having no spanwise domain boundaries. The width and height of the subset are $8\sqrt[3]{4}D_\mathrm{p}$, as in the flow simulations. We also remove the restriction on the inter-particle distance with and without spanwise domain boundaries. The best fit for the different combinations of restrictions are given in \cref{tab:AB}. \Cref{fig:sight-length-contours} shows a contour plot of the resulting sight-lengths for the range of particle volume fractions and particle diameters we use in the flow simulations. The figure also shows the contour lines of \cref{eq:sight-length} with the model constants when all particle distribution restrictions are applied. The model fits the data with a maximal relative error of $1.5\%$, and the maximal errors occur for low particle volume fraction and low particle diameter. Within the range of volume fractions we have simulated, the longest sight length is obtained with spanwise boundaries and no restriction on inter-particle distance. Compared to the case without restrictions, the case with spanwise boundaries has $17\%$ longer sight-length at $\alpha_\mathrm{p}=0.05$ and $15\%$ longer at $\alpha_\mathrm{p}=0.1$.
The inter-particle distance changes the sight-length with just $1\%$ at both ends of the volume fraction range. When both restrictions are combined, the sight-length is increased by $13\%$ at $\alpha_\mathrm{p}=0.05$ and by less than $1\%$ at $\alpha_\mathrm{p}=0.1$. Overall, the effect of the slightly increased regularity of the particle distribution is to increase the sight-length, but the effect is not very strong. When the particle volume fraction is doubled, the sight-length is almost halved. Therefore, variation of the sight length with changes in particle volume fraction is much more important than the variation with particle distribution restrictions used here. 

\subsection{Modeling the streamwise Reynolds stress}
\label{sec:r00model}
In this section we propose a model for the streamwise Reynolds stress based on the observations made in \cref{sec:momentum-balance}. Most importantly, we found that the Reynolds stress is caused mainly by separated flow behind the particles. 

\begin{figure*}
	\centerline{
		\includegraphics[]{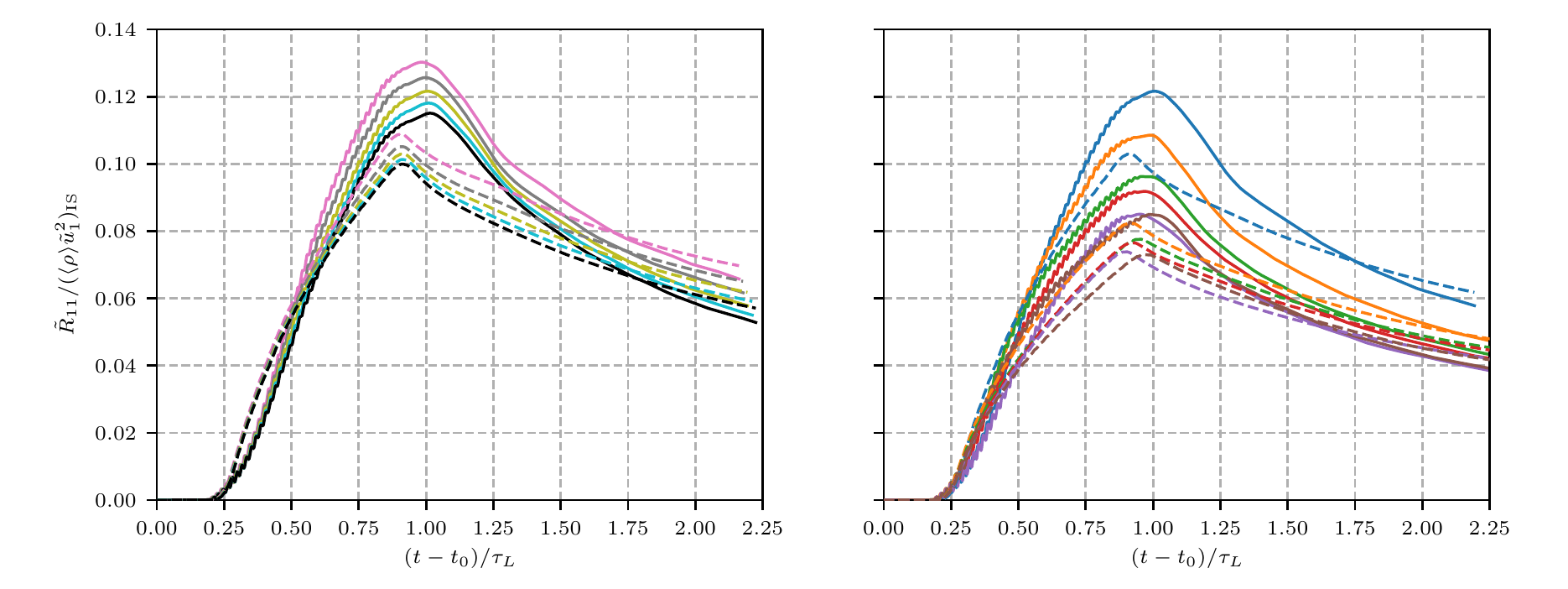}}
	\caption{Normalized streamwise Reynolds stress averaged over $0\leq x/L \leq 0.8$, as a function of time for the simulations with $\alpha_\mathrm{p}=0.1$, $D_\mathrm{p}=100\ \mu\mathrm{m}$ (left) and $Ma=2.6$ and various geometries (right). Line colors are as in \cref{fig:M2P6-uf,fig:Machs-uf}. Dashed lines are the corresponding fluctuations computed from \cref{eq:uumodel}.}
	\label{fig:wakemodel-average}
\end{figure*}

The gas within a separation region behind a particle has a velocity about the same as the particle, which in our case is zero. This is a few hundred m/s lower than the mean flow speed, and therefore the contribution to $\tilde{R}_{11}$ by separated flow regions will quickly become very large with increasing volume of the separation region. A simple model for this effect is the following: We assume that the separation region can be approximated as a region with velocity equal to the particle velocity. The rest of the volume then contains gas with an average velocity that is higher than $\favg{u}$, and related to $\favg{u}$ by a function of the volume of the separated flow. The sum of the volume of separated flow regions within a bin occupies a fraction of the volume which we denote by $\alpha_\mathrm{sep}$. If the mean velocity in the bin is known, in absence of any other sources of Reynolds stress, the streamwise velocity correlation is given as
\begin{equation}
\langle u_1''u_1'' \rangle = \favg{u}_1^2\frac{\alpha_\mathrm{sep}}{\alpha-\alpha_\mathrm{sep}}.
\label{eq:uumodel}
\end{equation}
Here, $\alpha_\mathrm{sep}$ is the volume fraction of the separation regions, which is a function that must be determined. These assumptions also imply that the mean velocity in the rest of the gas volume is
\begin{equation}
u_\mathrm{free} = \favg{u}_1\frac{\alpha}{\alpha-\alpha_\mathrm{sep}}.
\label{eq:ufree}
\end{equation}
We further assume that the triple-correlation part of	
\begin{equation}
\tilde{R}_{11}=\phavg{u''_1u''_1} + \phavg{\rho'u''_1u''_1}/\phavg{\rho},
\end{equation}
is negligible, in line with the principle of receding influence, so that $\favg{R}_{11}$ is directly given by \cref{eq:uumodel}. We evaluate $\alpha_\mathrm{sep}$ for the various simulations at every location between $0.2 \leq x/L \leq 0.8$, for the time interval between $15\tau_\mathrm{p}$ after the shock has passed that location until the end of the simulation. This time-delay is introduced so that the shock-induced transient does not affect the results. The average separation volume fraction for each simulation is given in \cref{tab:alpha_sep}. It was found to be insensitive to the location within $0.2\leq x/L \leq 0.8$, and insensitive to the incident shock wave Mach number. It varies slightly with particle diameter and drastically with volume fraction, which is to be expected. The ratio $\alpha_\mathrm{sep}/\alpha_\mathrm{p}$ increases significantly as $\alpha_\mathrm{p}$ is reduced. This is likely caused by the increasing inter-particle distance, which allows the separation regions to develop with less disturbances than for higher particle volume fractions. \Cref{fig:wakemodel-average} shows the normalized Reynolds stress obtained in the simulations, averaged over $0.2 \leq x/L \leq 0.8$ as a function of time, and the Reynolds stress computed from \cref{eq:uumodel}. The model captures the magnitude and general trend of the Reynolds stress well. In each case, there is a transient during and slightly after the shock wave passes through the particle cloud that the model does not represent. During this time, the shock reflection from individual particles comprise a significant portion of the observed Reynolds stress. Since the Reynolds stress model represents the effect of particle wakes only, we expect a deficiency when shock-reflection is still occurring. \Cref{fig:wakemodel-average} does indeed show that between $0.5\leq(t-t_0)/\tau_L\leq1.5$ the observed Reynolds stresses are higher than the model predicts. However a large portion of the Reynolds stress is also captured during this phase, because particle wakes develop for a large number of the particles before the shock exits the particle cloud. The model also increases too rapidly early on, since it is directly proportional to the mean velocity, while the separation regions take some time to develop. This model apparently captures the majority of the Reynolds stress, but it does not represent flow structures that are not associated with the particles, such as vortices being advected by the flow. We thus expect that the volume fractions reported in \cref{tab:alpha_sep} are slightly too high. Additional models should be used to capture the other phenomena causing velocity fluctuations. The model represents the velocity fluctuations caused by the separated flow regions behind each particle, and the additional velocity "fluctuations" that appear since the averaged flow velocity is shifted away from the free flow speed.

As a Reynolds stress model to be used in Eulerian-Lagrangian simulations of shock particle interaction, the form of \cref{eq:uumodel} is attractive, since it can be associated with the Lagrangian particles. It does not involve any gradient operations or additional interpolations than what must be used to compute particle drag. However, in Eulerian-Lagrangian simulations, the model is limited to situations where the control volumes are considerably larger than a single particle.

\begin{table*}
		\begin{tabular}{ccc@{\hskip 1.25cm}ccc}
			Case & $(Ma,\ \alpha_\mathrm{p},\ D_\mathrm{p})$ & $\alpha_\mathrm{sep}$ & Case & $(Ma,\ \alpha_\mathrm{p},\ D_\mathrm{p})$ & $\alpha_\mathrm{sep}$ \\
			\hline
			I &$(2.2,\ 0.1,\ 100\ \mu\mathrm{m})$ & 0.152 & VII &$(2.6,\ 0.1,\ 63.0\ \mu\mathrm{m})$  & 0.142 \\
			II &$(2.4,\ 0.1,\ 100\ \mu\mathrm{m})$  & 0.152 & VIII &$(2.6,\ 0.075,\ 57.2\ \mu\mathrm{m})$ & 0.131 \\
			III &$(2.6,\ 0.1,\ 100\ \mu\mathrm{m})$ & 0.153 & IX & $(2.6,\ 0.05,\ 50\ \mu\mathrm{m})$   & 0.115 \\
			IV & $(2.8,\ 0.1,\ 100\ \mu\mathrm{m})$  & 0.153 & X & $(2.2,\ 0.1,\ 50\ \mu\mathrm{m})$    & 0.148 \\
			V & $(3.0,\ 0.1,\ 100\ \mu\mathrm{m})$   & 0.152 & XI & $(2.6,\ 0.1,\ 50\ \mu\mathrm{m})$   & 0.148 \\
			VI & $(2.6,\ 0.1,\ 79.4\ \mu\mathrm{m})$  & 0.152 & XI & $(3.0,\ 0.1,\ 50\ \mu\mathrm{m})$  & 0.148 \\
		\end{tabular}
	\caption{Volume fraction of separated flow, $\alpha_\mathrm{sep}$, for the various simulations in this study.}
	\label{tab:alpha_sep}
\end{table*}

The assumptions made about the flow in the immediate proximity of the particles in the derivation of \cref{eq:uumodel} has implications for the calculation of particle drag. The model for $\favg{R}_{11}$ suggested above showed that between $10$\% and $15$\% of the volume could be considered to belong to separation regions. We assumed that the average velocity of a separation region was zero. The large volume in these separated flow regions means that the average velocity is not the appropriate velocity to use for computing the drag. If $u_\mathrm{free}$ is used instead of $\favg{u}_1$ to compute the drag coefficient, it will be lower than if it is computed directly from $\tilde{u}_1$. The velocity correction factor is $\alpha/(\alpha-\alpha_\mathrm{sep})$, and since the denominator in \cref{eq:Cd} contains $\favg{u}_1^2$, the drag coefficient becomes about 30\% lower when we include this correction. The model also implies a correction to the free flow Mach number, since the Mach number is proportional to $\favg{u}_1$. Standard drag laws correlate the particle forces to the undisturbed flow quantities, but since the notion of an undisturbed flow is meaningless inside a particle cloud, it is still an open question how to calculate drag in Eulerian-Lagrangian models. However, this model does partly explain why the average drag coefficients appear to be very high in this flow.  

The Reynolds stress model, based on an assumed separated flow in the particle wake with a given volume and an average velocity of zero, approximates the bulk streamwise Reynolds stress well. It is easily applicable in both Eulerian-Lagrangian and Eulerian-Eulerian models of dispersed flow, and only needs an estimate of $\alpha_\mathrm{sep}$, since the other model inputs are already known. The model implies corrections to mean flow properties, due to the non-negligible volume fraction of the separated flow. It is clear that improvements to this model can be obtained by detailed examination of the statistical properties of particle wakes in the shock particle setting, and this is a possible direction for future work.

\section{Concluding remarks}
\label{sec:conclusions}

In this work, we have investigated the flow fields during the passage of a shock wave through a random array of particles, using viscous three-dimensional numerical simulations. The flow field variation with different incident shock wave Mach numbers, particle volume fractions and particle diameters was investigated using the volume averaging framework. It was found that many mean flow features could be predicted based on the sight-length, which relates particle volume fractions and particle diameters. Flows at a given particle volume fraction were found to have many similarities to higher or lower particle volume fractions if the particle diameter was increased or decreased so that the sight-length remained the same. This characterization was found to work best within the interior of the particle layer. At the upstream and downstream particle cloud edges, we found that flow field fluctuations play an important role in the mean flow dynamics. The variation of the fluctuations was not found to be predicted by the sight-length parameter. In agreement with this, we observed that the quantities that varied predictably with sight-length in the interior of the particle cloud did not follow the same variation at the edges.

The regularity of the particle distribution has a non-negligible effect on the flow. This work utilized structured meshes around each particle, in order to best facilitate the prediction of the viscous shear layers. For this reason, a slightly increased regularity of the particle distribution was necessary. In the cases considered in this work, the regularity increased the sight-length. However, the change in sight-length due to regularity was found to be small compared to the change when the particle volume fraction was varied. In general, the specific realization of any particle distribution plays a role in the flow statistics, unless the number of particles in the spanwise directions is very large. Ensemble averaging can serve as a substitute for large computational domains, and is therefore recommended for future works. 

We observed that the flow expansion at the downstream edge of the particle cloud featured an increase in the mean free path of the air molecules. For the case with a $Ma=2.6$ incident shock wave, $\alpha_\mathrm{p}=0.1$ and $D_\mathrm{p}=63\ \mu\mathrm{m}$, we found that the Knudsen number of the local shear flow attained values above one around the particles furthest downstream. Different combinations of volume fractions and particle sizes might result in situations where non-continuum effects become significant, and should be treated carefully.

The Reynolds stress was found to be dynamically important at the particle cloud edges. The magnitudes of the Reynolds stresses were significantly higher than those found in the inviscid simulations of \citet{mehta2018-b}. Since the flow dynamics around the upstream particle cloud edge affect the reflected shock, this implies that the reflected shock is strengthened due to viscous effects. Around the particle cloud edges, the Reynolds stress gradients were of the same order as the particle forces. The Reynolds stress is caused primarily by shock wave reflection and separated flow behind the particles. It contains a significant part of the kinetic energy of the flow, and it is strongly anisotropic. Based on which physical phenomena that are the causes of the Reynolds stress, we proposed an algebraic Reynolds stress model. It uses the mean flow speed and the particle volume fraction, in addition to an estimate of the volume fraction of separated flow. The model could predict the magnitude of the Reynolds stress and its evolution in the interior of the particle cloud fairly well. By comparison to previous particle-resolved studies, the pseudo-turbulent kinetic energy magnitude was found to be more similar to results from incompressible simulations than to the results from previous shock wave particle cloud simulations. The previous shock wave particle cloud simulations were either inviscid or two-dimensional, and this has strong implications for the fluctuations within the particle cloud. We also found that the local Mach number was not very high within the particle layer after the shock-induced transient had ended. These observations suggest that the dynamics governing the flow fluctuations in the interior of the particle cloud are primarily incompressible flow phenomena.

\section*{Acknowledgments}
The authors acknowledge use of computational resources from the Certainty cluster awarded by the National Science Foundation to the Center for Turbulence Research, Stanford University.

\section*{References}
\bibliography{ShockParticleCloudBib.bib}

\end{document}